\documentclass[journal,transmag]{IEEEtran}

\usepackage{booktabs}
\usepackage{latexsym}
\usepackage{amsmath}
\usepackage{amssymb}
\usepackage{url}
\usepackage{ifpdf}
\ifpdf
  \usepackage{thumbpdf}
\else
\fi
\usepackage{cite}
\usepackage{color}
\usepackage[table]{xcolor}    
\usepackage{colortbl}
\usepackage{balance}
\usepackage{soul}

\usepackage{float}
\usepackage{epsfig}
\usepackage{epstopdf}
\usepackage[justification=centering]{caption}
\usepackage{graphicx}
\usepackage{multirow}
\usepackage{subcaption}
\usepackage[font={small}]{caption}
\usepackage{cuted}
\usepackage[lined,boxed,commentsnumbered,ruled,vlined,noend]{algorithm2e}

\IEEEoverridecommandlockouts

\begin{document}

\title{M2M Traffic via Random Access Satellite links:\\Interactions between Transport and MAC Layers}

\author{
  Manlio Bacco, Tomaso De Cola, Giovanni Giambene, Alberto Gotta 
  \thanks{(Corresponding author: Manlio Bacco, manlio.bacco@isti.cnr.it)}
  \thanks{Manlio Bacco and Giovanni Giambene are with the Department
  of Information Engineering and Mathematic Science, University of Siena, (e-mail: giambene@unisi.it).}
  \thanks{Manlio Bacco and Alberto Gotta are with the Information Science and Technologies Institute (ISTI), National Research Council (CNR), 56124, Pisa, Italy, (e-mail: alberto.gotta@isti.cnr.it)}
  \thanks{Tomaso De Cola is with the German Aerospace Center (DLR), Institute of Communications and Navigation, 82234, Oberpfaffenhofen, Germany, (e-mail: tomaso.decola@dlr.de)}
}


    
    
    

\maketitle

\begin{abstract}
\textit{Abstract} Machine-to-machine services are witnessing an unprecedented diffusion, which is expected to result in an ever-increasing data traffic load. In this context, satellite technology is playing a pivotal role, since it enables a widespread provisioning of machine-to-machine services. In particular, oil industry, maritime communications, as well as remote monitoring are sectors where the use of satellite communications is expected to dramatically explode within the next few years. In the light of this sudden increase of machine-to-machine data transported over satellite, a more thorough understanding of machine-to-machine service implementation over satellite is required, especially focusing on the interaction between transport and MAC layers of the protocol stack. 
Starting from these observations, this paper thoroughly analyses the interaction between TCP and the Contention Resolution Diversity Slotted Aloha access scheme defined in the DVB-RCS2 standard, assuming the use of an MQTT-like protocol to distribute machine-to-machine services. A novel TCP model is developed and validated through extensive simulation campaigns, which also shed important lights on the design choices enabling the efficient transport of machine-to-machine data via satellite. 
\end{abstract}


\section{Introduction}
\label{sec:ton_introduction}

According to data traffic forecast reports, the volume of data transported by Internet in 2019 will exceed the threshold of 2.0 zettabytes per year, generated by more than trillion of devices. Only a minor portion of the traffic will be generated by PCs, as commonly observed in the recent past; on the contrary, a large quota of Internet traffic will be generated by TVs, tablets, smartphones, and Machine-to-Machine (M2M) devices. In particular, it has been highlighted that M2M traffic will experience a growth rate in the order of 60\%. The large amount of traffic contributed by M2M applications will have an important impact on the design of future network architectures and on dimensioning the capacity of the telecommunication infrastructures. 
In more words, M2M applications are today largely diffused in several terrestrial deployments and have pushed the scientific community to thoroughly investigate network design implications. The survey in \cite{kim2014m2m} underlines that current M2M markets are fragmented, because various vertical M2M solutions have been already designed and implemented: the efforts to identify a common architecture is of a primary concern. Moreover, reference \cite{kim2014m2m} also shows that typical Internet of Things (IoT)/M2M services exhibit intermittent behaviour (e.g., ON-OFF oscillations), low data-rates, and high traffic burstiness, thus motivating the use of Random Access (RA) schemes. Seminal works about the use of RA for the efficient transport of M2M services via satellite are provided in \cite{herrero2012high,scalise2013s}.
The use of satellites is expected to become crucial in the near future \cite{cerwall2015ericsson, idate2010forum}, because very highly dense networks, such as M2M ones, pose several challenges to terrestrial wireless networks \cite{laya2014random}. 
In fact, the availability of satellite connectivity is of paramount importance in scenarios such as remote plant monitoring, ship tracking service, and aeronautical telemetry transmission, just to mention a few, where a typical terrestrial infrastructure cannot be exploited. 
From this standpoint, a special note has to be reserved to the case of M2M services distributed via satellite, whose related industry is continuously increasing in size. 

The design of efficient RA schemes for satellite communications has captured a great interest from the scientific community and the satellite industry over the last decade \cite{herrero2012high,scalise2013s}. An important example is represented by the second generation of DVB-RCS (DVB-RCS2) \cite{dvbrcs2}, specifying the transmission schemes and the protocol architecture for the return link, where Contention Resolution Diversity Slotted Aloha (CRDSA) \cite{crdsa} is proposed as a complementary method to the traditional Demand Assignment Multiple Access (DAMA). CRDSA offers immediate access to satellite capacity without incurring in the reservation delays of DAMA that can penalise the transmission of bursty traffic (e.g., M2M). 
CRDSA consists in a more sophisticated version of Slotted Aloha (SA), where multiple replicas of the same MAC packet are sent and where accidental collisions are solved by means of Successive Interference Cancellation (SIC). CRDSA relies on the use of two replicas, while the most recent CRDSA++ protocol considers also the use of more than two replicas \cite{del2009high}. 

The focus of the scientific community has been so far mostly around the implications of RA on physical layer design, only partly addressing its interactions with upper layers (e.g., TCP/IP stack). In particular, the performance analysis of TCP-based higher layer protocols running on top of RA schemes has not yet provided meaningful indications about the boundaries of its applicability, apart from the contribution in \cite{collina2014internet}, which anyway offers a simple and preliminary empirical evaluation.
Furthermore, the impact of the collision resolution scheme on TCP congestion control in terms of delay  as well as the reaction of TCP in response to TCP segments' collision still needs a rigorous investigation from a theoretical standpoint.
This paper attempts to bridge this scientific gap, by analyzing the performance of TCP-based application protocols over CRDSA++, using 3 replicas and running in a DVB-S2/RCS2 network under different traffic load conditions. To this end, a rigorous analytical model of TCP dynamics over RA is drawn, adapting and extending the TCP models in \cite{padhye2000modeling, parvez2010analytic, dunaytsev2007analytical} that cannot inherently apply to the case of RA schemes. The analytical approach will be validated via NS-3-based simulations \cite{ns3}.

The remainder of this paper is structured as follows: Section \ref{sec:ton_overview} surveys the main findings on TCP modeling and interactions with RA schemes, paying special attention to the case of satellite networks. Section \ref{sec:ton_architecture} introduces the reference system, while Section \ref{sec:ton_model} proposes some refinements that are necessary on long-delay RA links in order to correctly estimate TCP throughput. The performance analysis is carried out in Section \ref{sec:ton_performance}. Section \ref{sec:ton_pl} offers a simple but effective model to link together burst losses at MAC layer, due to collisions, and segment losses at transport layer. Finally, conclusions are drawn in Section \ref{sec:ton_conclusions}.
\section{Overview}
\label{sec:ton_overview}
\subsection{Related works}
\label{subsec:ton_related}

In this Section, a survey is presented about the theoretical models available in the literature for estimating the TCP throughput. 
The literature about TCP modeling can be divided into two main sets, characterized by different packet loss models: bursty losses \cite{Cardwell00modelingtcp,padhye2000modeling} and independent losses\footnote{As far as contention-based medium access schemes (e.g., CRDSA) are concerned, losses due to collisions should be considered as independent events: this is empirically proved in Section \ref{sec:ton_delta}.} \cite{1255433,ZhouYeung}. The TCP behavior is modeled as a \textit{renewal process} for the first time in \cite{padhye2000modeling} and then in \cite{ZhouYeung}. 
A TCP \textit{cycle} is defined as the period of time between to consecutive loss events and it is equivalent to an \textit{epoch} of the corresponding renewal process. The performance of TCP is modeled in terms of \textit{rounds}: an epoch contains $m$ rounds. A round begins with the transmission of \textit{W} segments, called Congestion Window (CWND), and ends upon the receipt of the Acknowledgments (ACKs) corresponding to these segments.
The duration of a round is assumed to be independent of the CWND size and dependent on the Round Trip Time (RTT). The CWND grows at each round, until a loss event occurs; then, TCP enters the recovery phase. Apart from the recovery algorithm that may vary according to the TCP flavor, TCP leaves the recovery phase with or without the expiration of the TimeOut (TO), thus entering a new epoch. The first rounds of the new epoch correspond to either a Slow Start (SS) or a Congestion Avoidance (CA) phase, the former if TO has expired, the latter otherwise.
In \cite{padhye2000modeling}, a Triple-Duplicate ACK (TD) is used as loss indication and a TD Period (TDP) (i.e., the period between two TD loss indications) coincides with an epoch of the renewal process.
Since our aim is to establish a relationship between the throughput $T$ of a TCP connection and the segment loss probability $p$, let $S_i$ be the number of TCP segments sent during the $i$-th epoch and $D_i$ the duration of that period. Thus, we can express the TCP throughput as follows:
\begin{equation}  
  T=\frac{\mathbb{E}[S_i]}{\mathbb{E}[D_i]}.
  \label{eq:ton_thr}
\end{equation}

The main outcomes in \cite{padhye2000modeling}, then refined in \cite{towcomments}, are the so-called Square Root Formula (SRF) and the Approximated Model (AM). The model in \cite{padhye2000modeling} treats segment losses as independent events in different rounds and as correlated events in the same round: each segment delivered after the first segment loss is supposed to be lost as well, independently of its actual outcome. The throughput models in \cite{padhye2000modeling} have been derived for TCP Reno recovery algorithm, which can recover a single loss during the Fast Retransmit / Fast Recovery (FR)\footnote{RFC 6582: The NewReno Modification to TCP's Fast Recovery Algorithm.}. However, more recent TCP implementations, like NewReno or Cubic, have more complex error recovery mechanisms that allow recovering multiple losses during the FR phase. Other works have extended SRF and AM in order to account for FR phase' dynamics with Slow But Steady (SBS) flavor \cite{parvez2010analytic, dunaytsev2007analytical}. 
The most prominent innovation in \cite{parvez2010analytic} and \cite{dunaytsev2007analytical}, with respect to \cite{padhye2000modeling}, is that both bursty and independent losses are modelled.

When specifically targeting the use of TCP in RA schemes for satellites, the work in \cite{liumodiano} provides upper and lower bounds of the throughput for a simplified TCP model in a Slotted Aloha (SA) RA channel. Furthermore, the TCP performance over a satellite Diversity Slotted Aloha (DSA) \cite{choudhury1983diversity} RA medium is analyzed for a single TCP connection in \cite{Celandroni_2014}; the authors assume that the MAC layer operates at a target load $G^*$ and that the Network Control Center (NCC) periodically broadcast an \textit{activity factor} (i.e., a throttling parameter) to dynamically adjust the proposed flow control policy applied at Return Channel Satellite Terminals (RCSTs), in order to maximize the TCP throughput. Analogously, \cite{celandroni2014elastic} analyzes TCP throughput over CRDSA when packet-level Forward Error Correction (FEC) techniques are employed; instead, in this work, we do not use any packet-level FEC techniques. Furthermore, a single TCP connection is modeled through a fluidic model in \cite{celandroni2014elastic}, less precise than the packet level simulator in use in this work. 
Finally, reference \cite{bacco2015advances} considers a similar scenario, i.e. the use of MQTT in RA satellite channels, but it focuses on a different metric, that is the \textit{completion time}. In fact, the broker is placed close to the subscribers, thus each publisher delivers its data via satellite to the remote broker. Furthermore, a simplified scenario is taken into account: low/medium traffic conditions are under consideration and the impact of packet fragmentation is not analyzed.

\subsection{Main contributions}
\label{subsec:ton_contribution}
This work extends the contribution in \cite{bacco2015tcp}, where the problem was firstly stated and discussed. Here, a complete set of MAC parameters is taken into account in the analysis to better describe the behavior of TCP over a CRDSA-based satellite channel. This work also proposes to use TCP for sending IoT/M2M traffic over CRDSA++, removing the need for any other flow control scheme
in use at MAC layer.

To summarize, the main contributions in this work are:
\begin{itemize}
\itemsep0em 
\item the analysis of how the TCP congestion control algorithm behaves in a channel dominated by collisions, where TCP auto-regulates the sending rate to counteract an increase in Segment Loss Rate (SLR); 
\item a finer model derived from \cite{padhye2000modeling, parvez2010analytic, dunaytsev2007analytical}, which accurately matches the TCP steady state throughput over RA satellite channels;
\item a simple but effective model establishing a relation between the loss rate at MAC layer and the loss rate at transport layer (cross-layer study);
\item finally, considerations on system stability guaranteed by the TCP control congestion algorithm, even at very high loads.
\end{itemize}
\section{System Architecture and Setup Parameters}
\label{sec:ton_architecture}

\subsection{Reference scenario}
\label{subsec:ton_scenario}
\begin{figure}
  	\centering
  	\includegraphics[scale=0.45, clip=true, trim=0 0 0 0]{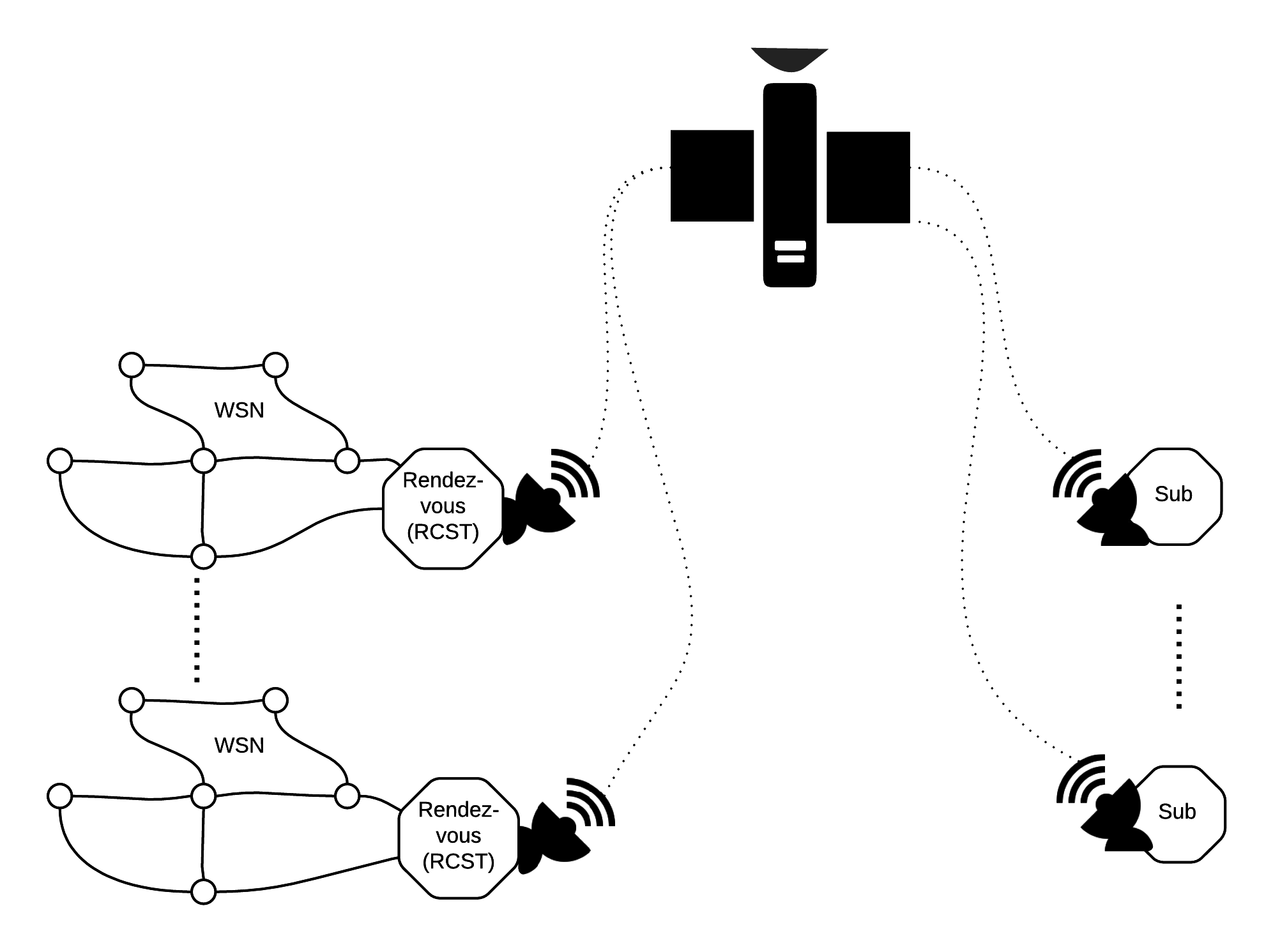}
	\caption{Scenario under investigation: multiple RCSTs acting as rendezvous nodes and delivering data to remote subscribers via RA satellite link}
  	\label{fig:ton_net1}
\end{figure}

This work focuses on the applicability of the DVB-RCS2 standard to scenarios in which IoT/M2M traffic is generated from Wireless Sensor Networks (WSNs), similarly to what is considered in \cite{desatellite}. The collected data are \textit{published}, according to the Publish/Subscribe (PUB/SUB) paradigm; a \textit{broker} node collects data and makes them available to \textit{subscribers}.
In the scenario depicted in Figure \ref{fig:ton_net1}, each WSN has a local broker node (also referred to as \textit{cluster head} or \textit{rendezvous} node); the subscribers are connected via a satellite link. Several WSNs are deployed on the ground and each WSN has a potentially large number of sensors, which periodically generate IoT/M2M data. 
Each broker has no constraints about power consumption and available computational resources; it is connected to a RCST, sending collected data via an RA satellite channel. 

The broker is responsible for the delivery of data collected by the WSN. The broker is supposed to have an endless queue of IoT/M2M data waiting for transmission, so that the offered traffic is sustained.
Such assumption is valid in several possible application scenarios: for instance, in battle-field communications, where several disjoint teams can use several WSNs sending and receiving data from a remote command center; in maritime scenarios for fleet management; in Flying Ad-Hoc Networks (FANETs), where several resource-constrained Unmanned Aerial Vehicles (UAVs), composing a swarm, can rely on a local rendezvous node or acting as peers to transmit data to a remote rendezvous node. 
In these scenarios, we will provide an analytical model in Sections \ref{sec:ton_model} and \ref{sec:ton_pl} to predict TCP throughput in supporting IoT/M2M traffic over a satellite RA channel affected by segment losses due to collisions.

According to ETSI (European Telecommunications Standards Institute) and IETF (Internet Engineering Task Force), the Constrained Application Protocol (CoAP)\footnote{IETF RFC 7252.} is the reference protocol for M2M\footnote{ETSI 102 690 technical specifications indicate CoAP as reference protocol for M2M scenarios.}. Another IoT/M2M protocol, MQTT\footnote{MQTT v3.1 protocol specification, IBM developerWorks Technical Library, 2010.}, developed by IBM in 1999, is the main opponent to CoAP, with a large history of successful deployments. 
The two protocols have several differences. For instance: (\textit{i}) MQTT is TCP-based, instead CoAP is UDP-based; (\textit{ii}) MQTT has (virtually) no limits on the payload size, while CoAP is limited to 1280 bytes; (\textit{iii}) MQTT is based on a PUB/SUB paradigm, while CoAP provides a request/response interaction model and is REST-oriented (Representational State Transfer). CoAP has been designed with constrained devices in mind, and this explains the lightweight transport protocol in use and the limitation on the payload size, among others. 

MQTT has been selected as reference protocol in this work because it is more common in real installations. Furthermore, the PUB/SUB paradigm implemented by MQTT foresees the use of a broker node (as already considered in the architecture in Figure \ref{fig:ton_net1}), which acts similarly to a Split-TCP PEP (Performance Enhancement Proxy), widely used in satellite networks. 

We suppose that no data is lost (reliable delivery) in the local WSN among the IoT/M2M nodes and the local rendezvous node. The subscribers receive data from the rendezvous node via an RA satellite link.
An MQTT-like application TCP-based protocol is assumed at the application layer, ensuring a reliable delivery of data from rendezvous nodes to subscribers. The only different assumption is that the underlining TCP connection is open only at the very first data transmission, to avoid a continuous three-way handshake (3WHS) between a rendezvous node and one or more subscribers for the transmission of each data burst.
The performance metric under consideration is the throughput at transport layer that corresponds to the throughput experienced by the application itself, apart from the overhead introduced at the application layer. 
The coexistence of TCP-based and non-TCP-based traffic in the satellite return link is out of the scope of this work and is left for future studies.

\subsection{Satellite system}
\label{subsec:ton_satellite}
Let us consider a satellite system and a finite number $N$ of RCSTs, sending data via a satellite RA return link. The RA protocol is CRDSA++ \cite{del2009high}.
Time is slotted and each superframe contains one or more frames; according to DVB-RCS2 terminology, a frame on a return channel is also referred to as RA block. In our configuration, an RA block spans the whole superframe and is composed of $n$ time-slots, or \textit{transmission opportunities}. Each RCST can exploit any transmission opportunity to send data, also referred to as \textit{bursts}, over the return channel. In this work, each transmission opportunity can be used with probability equal to 1, if any data is ready for transmission.
At layer 2, the Return Link Encapsulation (RLE) protocol encapsulates the datagrams coming from the network layer. RLE handles \textit{encapsulation}, \textit{fragmentation} and \textit{frame packing} procedures. According to the upper-layer datagrams size, an RLE data unit can carry: a fraction of a datagram, the whole datagram or more than one datagram, according to the respective lengths. 
Let $r$ denote the length of an RLE data unit measured in IP datagrams that can be packet together, while $f \geq 1$ is defined as the number of IP datagrams that are lost in a single MAC-layer collision event on the RA channel. In case of a collision, $f=\lceil r \rceil$ IP datagrams must be retransmitted. We assume no IP fragmentation in the following, so that one IP datagram carries a whole TCP segment.

We assume that the RCSTs are power balanced\footnote{In the reference scenario, RCSTs are uniformly distributed within the same beam. We assume that an uplink power control scheme is implemented, so that the power of the signal received from each RCST is approximately the same.}.
It is worth remarking that power-balancing is the worst condition RCSTs can face when using RA schemes, because the power unbalancing improves the benefits offered by the \textit{capture effect}, thus improving the performance of SIC.

The DVB-RCS2 standard defines two burst lengths ($536$ and $1616$ symbols, respectively) to transport user data that can be transmitted by means of a rich set of waveforms (WFs), so as to operate in different channel conditions and to support various traffic types\footnote{Note that DVB-RCS2 also defines two additional burst lengths ($664$ and $262$ symbols, respectively) transported by WF 1 and WF 2, which are however used only for control messages. Hence, they are not considered in this work.}. In more detail, the redundancy added by channel coding schemes as well as the modulation order can be tuned according to the channel conditions: DVB-RCS2 provides several predefined configurations, or WFs, in order to satisfy reference Bit Error Rate (BER) figures. As a result, the time-slots of an RA block can accommodate different amount of data: a burst is defined as the amount of data fitting a time-slot. This feature can be exploited to configure DVB-RCS2 devices so as to handle different traffic types, such as small-payload messages characterising M2M-like traffic, or conversely larger ones. 
In the present work, WF 14 and WF 3 configurations are selected (see Table \ref{table:ton_wf_details} for more details about these WFs). Apart from the chosen WFs, all the other parameters are equal in the scenario under consideration. The RA block of WF 14 is composed of time-slots, each carrying 188 bytes; the RA block of WF 3 is composed of time-slots, each carrying 38 bytes. Note that WF 3 is one of the most robust and used waveforms, relying on Quadrature Phase Shift-Keying (QPSK) $1/3$, while WF 14 uses QPSK $1/2$. The choice of these two WFs stems from the fact that the corresponding volume of data being transported can be considered as representative of typical M2M message lengths. 
\begin{table*}
\caption{Details on the DVB-RCS2 waveforms in use}
  \begin{center}
  \begin{tabular}{|c|c|c|c|c|} 
  \hline   
  \textbf{Waveform ID} & \textbf{Burst length [symbols]} & \textbf{Payload length ([B], [symbols])} & \textbf{Mapping scheme} & \textbf{Code rate} 	\\	  \hline \hline
  3 & 536 & 38, 456 & QPSK & 1/3					\\ 	\hline
  14 & 1616 & 188, 1504 & QPSK & 1/2				\\ 	\hline    
  \end{tabular}
  \end{center}
  \label{table:ton_wf_details}
\end{table*}

\subsection{TCP segment size and fragmentation at MAC layer}
\label{sec:ton_mss}
Two different Maximum Segment Sizes (MSSs) are under consideration: 23 and 173 bytes, meant to exactly fit the payload size provided by WF 3 and by WF 14, respectively. IoT/M2M terminals generate low data-rates, thus the MSSs here considered are enough to periodically transmit small amount of data. 
Since we are dealing with small IoT/M2M payloads, RObust Header Compression (ROHC) is used to reduce the impact of TCP/IP overhead. ROCH is applied prior to transmitting data on the satellite return link. The payload size of the data stream is assumed to be constant, thus the 52 bytes of TCP/IP headers\footnote{40 bytes for TCP/IP headers plus 12 bytes for TCP options.} reduce to 7 bytes for DATA segments and to 6 bytes for ACK segments\footnote{See Section 4.4 of RFC 6846.}. The compression ratio for DATA is larger than 85\%, thus effectively reducing the overhead. 

Several scenarios are possible, depending on the chosen MSS and WF, because of the fragmentation that can take place at MAC layer prior transmitting a TCP segment, as described in Table \ref{table:ton_fragmentation}, where we consider that 15 bytes of headers (comprised TCP/IP headers plus layer-2 overhead) have to be summed to MSS.
Three scenarios are possible, depending on the time-slot payload size:
\begin{enumerate}
\itemsep0em 
\item the MSS has the same length ($r=1$): one TCP segment exactly fits one MAC burst, thus a collision event causes a single segment loss, i.e., $f=1$;
\item the MSS has a larger length ($r<1$): one TCP segment fits into several MAC bursts. A collision event causes a single segment loss, because losing a fraction of the segment is equivalent to lose the whole segment, i.e., $f=1$;
\item the MSS has a smaller length ($r>1$): several segments fit into a single MAC burst, thus a collision event causes $f = \lceil r \rceil > 1$ segment losses.
\end{enumerate}
The latter case is discussed separately because it violates the hypothesis of uncorrelated losses in Section \ref{sec:ton_model}. 
Moreover, we will show in Section \ref{sec:ton_performance} that this configuration has a poor performance, so that any configurations leading to $f>1$ should be avoided, to reduce the possible waste of costly resources.

\begin{table}
\caption{Fragmentation at MAC layer}
  \begin{center}
  \begin{tabular}{|c|c|c|c|} 
  \hline  
 
  \textbf{MSS + headers [bytes]} & \textbf{WF ID} & \textbf{$r = \frac{RLE \thinspace payl.(WF)}{MSS + head.}$} & \textbf{$f = \lceil r \rceil$} \\	  \hline \hline
  23 + 15 & 3 & 1 & 1 		\\ 	\hline
  23 + 15 & 14 & 5.7 & 6 	\\ 	\hline
  173 + 15 & 3 & 0.175 & 1 	\\ 	\hline
  173 + 15 & 14 & 1 & 1 	\\ 	\hline
  
  \end{tabular}
  \end{center}
  \label{table:ton_fragmentation}
\end{table}

\subsection{Interactions between TCP and the underlying RA scheme}
\label{sec:ton_tcp}
The traffic arrival process for each RCST is driven by a sustained and continuous load at application layer, because each rendezvous node continuously collects IoT/M2M data from the WSN and sends them to the remote subscribers. 
The DVB-RCS2 standard specifies a layer-2 load control mechanism (see Section \ref{sec:ton_reg}) aimed at keeping the system running at a target operating point $G_T$. However, in this work, the actual operating point $G$ is determined by the TCP congestion control algorithm because it already provides a load control mechanism, even if it is meant for avoiding congestion events and not for dealing with RA collisions; because of this, the DVB-RCS2 load control algorithm is not adopted. Every time a collision event causes a burst loss, $f$ TCP segments are lost, leading to the retransmission of the segments and to a reduction of the TCP sending rate. 

TCP NewReno deals with segment losses relying on: (\textit{i}) the use of the FR mechanism; (\textit{ii}) the TO event, triggered if FR is not successful or entering FR is not possible. The CWND is, then, set to a lower value and the Slow Start Threshold (SST) is set accordingly. In both cases, the sending rate is reduced to counteract further loss events.
In this work, the satellite return/forward links are supposed operating in clear-sky conditions\footnote{DVB-RCS2 offers adaptive MODCOD (Modulation and Coding) to fulfill a target BER requirement. A quasi-error-free link assumption can be made by carefully choosing the target BER.} in order to focus only on segment loss events caused by collisions 
(ACKs are assumed to be always correctly received). No Automatic Repeat-reQuest (ARQ) algorithm is in use at MAC layer. Each TCP segment sent to the lower layers (i.e., network and MAC layers) is queued into a large finite buffer, so that buffer losses can be neglected w.r.t. collision losses.

This paper analyzes the capability of TCP to self-regulate its transmission rate; given $N$ TCP flows belonging to $N$ distinct RCSTs, the resulting offered MAC load per RCST is given by $\lambda = G/N$, i.e., the TCP flows contribute in achieving the maximum throughput at a load level $G$ that depends on the dynamics of the congestion control algorithm: this aspect will be deepened in Section \ref{sec:ton_reg}. 
\section{Analytical Model of TCP NewReno}
\label{sec:ton_model}
This Section presents some analytical refinements to the throughput estimation model for TCP NewReno on top of RA.
The throughput estimation models anticipated in Section \ref{subsec:ton_related} do not accurately match the empirical results of the simulation campaign, as shown in Section \ref{sec:ton_performance}, thus motivating the refinements resented here. We stress that the throughput estimation model proposed in this Section is meant for TCP over an RA satellite link, and it has not been tested under different conditions. In this analysis, no assumptions are made on the specific RA protocol in use.
Furthermore, given the IoT/M2M traffic profile under consideration, different TCP flavors may show basically the same performance level, because of limited CWND size and reduced available bandwidth, the latter being a typical case of satellite RA links.

The new model presented in the following, namely \textit{NewRenoSAT}, is firstly developed in Section \ref{subsec:ton_noTO} assuming that all loss events are identified by TD ACKs, and then in Section \ref{subsec:ton_Full} handling both TD ACKs and TOs (the so-called \textit{full model}).

\subsection{Throughput model if loss indications are TD ACKs only}
\label{subsec:ton_noTO}
In this Section, we derive the model for estimating the TCP throughput in absence of TOs, thus only considering Congestion Avoidance (CA) and FR phases.

The throughput $T$ of a TCP flow can be estimated by analyzing a TCP cycle, an epoch of the renewal process. Let $S_i^{CA}$ and $S_i^{FR}$ be the number of segments successfully transmitted during the CA and FR phases of the $i$-th cycle, respectively, and $S_i^{CAFR}$ the number of segments sent during a whole cycle, here denoted as CAFR (Congestion Avoidance - Fast Retransmit / Fast Recovery):
\begin{equation}  
  S_i^{CAFR}=S_i^{CA}+S_i^{FR}.
  \label{eq:ton_segment}
\end{equation}
Let $D^{CA}_i$ and $D^{FR}_i$ denote the time duration of CA and FR periods and $D_i^{CAFR}$ the duration of the $i$-th CAFR cycle:
\begin{equation}  
  D_i^{CAFR}=D_i^{CA}+D_i^{FR}.
  \label{eq:ton_duration}
\end{equation}

The average throughput of a flow is given by (\ref{eq:ton_thr}). During the CA phase, the receiver sends one ACK every \textit{b} segments it receives (\textit{delayed ACK} feature\footnote{As RFC 1122 suggests, the amount of traffic from the receiver to the sender should be reduced by sending a single ACK every $b$ segments.}), causing the CWND to increase linearly with a slope of $1/b$ segments per round, until the first segment loss occurs. Let us denote by $\alpha_i$ the first lost segment in the $i$-th cycle and by $X_i$ the round where this loss occurs.
The total number of segments sent in $i$-th cycle is $S_i^{CAFR}=\alpha_i+\gamma_i$, where $\gamma_i$ is the number of segments sent between the first loss $\alpha_i$ and the last one in the drop window $W_i$. A \textit{drop window} is defined as a CWND where a loss event has occurred. $S_i^{CAFR}$ is defined as in Eq. (2) in \cite{parvez2010analytic}:
\begin{equation}  
  S^{CAFR} = \mathbb{E}[\alpha] + \mathbb{E}[\gamma],
  \label{eq:ton_segmentsent}
\end{equation}
where $S^{CAFR} = \mathbb{E}[S_i^{CAFR}]$.
The expected number of segments sent in a cycle, having $k$ rounds, up to $\alpha_i$, is given in \cite{dunaytsev2007analytical} as:
\begin{equation}  
  \mathbb{E}[\alpha]=\sum\limits_{k=1}^{\infty} k(1-p)^{k-1}p=\frac{1}{p},
  \label{eq:ton_averagealpha}
\end{equation}
where \textit{p} represents the average loss event rate. A \textit{loss event} is said to happen when one or more losses occur in a CWND, thus triggering the entering in FR (or triggering a TO, as in Section \ref{subsec:ton_Full}). The first segment loss in a CWND marks the beginning  of a \textit{loss event}.
The condition for entering the FR phase is to successfully deliver at least three segments in the drop window $W_i$. Hence, let us assume $\delta_i \geq 1$ 
losses with rate $q$ over the remaining $(W_i-3)$ segments
, where $q$ corresponds to the average SLR.
The number of losses $\delta_i$ follows a binomial distribution over a drop window; in particular, let us consider the probability distribution of $\delta_i$ losses over $(W_i-3)$ segments, conditioned on $\delta \geq 1$, as:
\begin{equation}  
	\begin{split}
    &Prob\{\delta = j \thinspace | \thinspace \delta \geq 1, W_i > 3\} = B[W_i - 3, j] \\
	& =\binom{W_i - 4}{j - 1} (1-q)^{W_i-3-j}q^{j-1}, j \in [1,W_i-3].
	\end{split}
    \label{eq:ton_binomial} 
\end{equation}
Analogously to (\ref{eq:ton_averagealpha}), the average number of segments sent between two consecutive losses is $1/q$. Therefore, if $\delta_i$ losses are assumed, the average number of segments among $\delta_i$ losses in the same CWND is $(\delta_i-1)/q$. The expected value of $\gamma$ can be calculated as follows:
\begin{equation}  
  \begin{split}
  \mathbb{E}[\gamma] &= \thinspace \mathbb{E} \left[ \sum\limits_{j=1}^{W_i-3} \frac{(j-1)}{q} \thinspace \left(q \thinspace B[W_i - 3, j] \thinspace \right)\right] \\
     & = q \thinspace (\mathbb{E}[W] - 4),
\end{split}
  \label{eq:ton_gamma}
\end{equation}
where $\mathbb{E}[W]$ is the expected value of the CWND size, under the assumption of steady state, and $(q \thinspace B[W_i - 3, j])$ is the joint probability of the first loss $\alpha_i$ and of the other $(j-1)$ losses in the drop window. Note that $\mathbb{E}[\gamma]$ is defined only in presence of the losses.  
If just a single segment loss occurs, then $\mathbb{E}[\gamma] = 0$; otherwise, if two or more segment losses occur, then $\mathbb{E}[\gamma] > 0$.
When $q \to 1$, $\mathbb{E}[\gamma] \to (\mathbb{E}[W]-4)$, i.e., the number of segments sent in a drop window after the first three segments allowing for TD ACKs and the first loss $\alpha$; conversely, when $q \to 0$, $\mathbb{E}[\gamma] \to 0$.

In order to derive the average number of rounds in a cycle, namely $\mathbb{E}[X]$, and the average CWND size, let us consider the evolution of CWND as a function of the number of rounds. 
According to \cite{dunaytsev2007analytical} and \cite{towcomments}, we have:
\begin{equation}  
  W_i= \frac{W_{i-1}}{2}+\frac{X_i}{b}-1,
  \label{eq:ton_wi}
\end{equation}
thus we can write the following relation, at regime, among average values:
\begin{equation}  
  \mathbb{E}[X]=b\left(\frac{\mathbb{E}[W]}{2}+1\right).
  \label{eq:ton_edix}
\end{equation}   

The expected number of segments sent during the CA phase is determined in \cite{towcomments} as:
\begin{equation}
  S^{CA} = \frac{\mathbb{E}[X]}{4} \thinspace 3 \thinspace \mathbb{E}[W] + \mathbb{E}[\beta],
  \label{eq:ton_sCA}
\end{equation}
where $\beta_i$ is the number of segments sent in the last round of the \textit{i}-th cycle. 
$\mathbb{E}[\beta] \approx \mathbb{E}[W]/2$ because we assume $\beta$ to be uniformly distributed in $[1,W_i-1]$.

The number of segments $S_{i,k}$ sent at each round $k$ of the \textit{i-th} FR cycle, considering the Partial Window Deflation\footnote{RFC 6582: The NewReno modification to TCP’s fast recovery algorithm.} mechanism, can be derived from \cite{joo} as:
\begin{equation}
	S_{i,k}=\max\Big(0,\:W_{i}/2-\delta_i+k-1\Big),\; 1 \leq k \leq \delta_i.
	\label{eq:ton_segFR}
\end{equation}
The entering in an FR phase is due to the reception of a TD. When a partial ACK is received, CWND is decreased by the amount of data acknowledged and increased of one segment size. Therefore $S_{i,k+1} = S_{i,k} + 1$. 
$S_{i,k}$ has a maximum value for $k=\delta_i$, then $\max(S_{i,k}) = S_{i,\delta_i} = W_i/2 - 1$.
Thus, the expected total number of segments sent in an FR cycle comes from (\ref{eq:ton_segFR}) as follows:
\begin{equation}
  \begin{split}
  S^{FR}&=\mathbb{E}\Big[\displaystyle\sum_{k=1}^{\delta_i} S_{i,k}\Big]\\
  &=
  \begin{cases}
          \frac{1}{2}(\mathbb{E}[\delta]\:\mathbb{E}[W]-\mathbb{E}[\delta]-\mathbb{E}[\delta]^2),& \text{if } \mathbb{E}[\delta]< \mathbb{E}[W] \\
  0& \text{if } \mathbb{E}[\delta] = \mathbb{E}[W].
  \end{cases}
  \end{split}
  \label{eq:ton_sFR}
\end{equation}
In the case of a single loss, i.e., $\delta_i=1$, $S^{FR}=\mathbb{E}[W]/2-1$. 
After the first loss, the remaining $(\delta_i-1)$ segments are lost over $(W_i-4)$ segments. Hence, we can write the following formula to characterize $\mathbb{E}[\delta]$, the expected value of the number of losses $\delta_i$: 
\begin{equation}
  \mathbb{E}[\delta] = \mathbb{E} \left[ \sum_{j=1}^{W_i-3} j \thinspace B[W_i - 3, j] \right] \approx 1+(\mathbb{E}[W]-4) \thinspace q.
  \label{eq:ton_delta}
\end{equation}
Assuming that collisions on an RA channel are independent events and $f=1$, then the segment losses are independent events, too. Instead, if $f > 1$, each collision in the channel causes a burst of segment losses of average length in the interval $[1,f]$, thus TCP segment losses cannot be considered as independent ones. Further details on this are in Section \ref{sec:ton_delta}.

Using (\ref{eq:ton_sCA}) and (\ref{eq:ton_sFR}), we can express the expected number of segments sent during a CAFR cycle, i.e., the expected value of (\ref{eq:ton_segment}), as follows:
\begin{equation}
  S^{CAFR} = S^{CA} + S^{FR}.
  \label{eq:ton_esi}
\end{equation}
Then, the expected value of CWND, $\mathbb{E}[W]$, can be obtained by equating (\ref{eq:ton_segmentsent}) and (\ref{eq:ton_esi}).
In particular, we obtain a second order equation in $\mathbb{E}[W]$, whose only positive solution can be expressed as:
	\begin{equation}
  	\mathbb{E}[W]\approx\Phi+\sqrt{\Phi^2-\frac{60pq-8p-8}{p(8q+3b)}},
  	\label{eq:ton_window}
	\end{equation}
where $\Phi=\dfrac{22q-3b-4}{8q+3b}$. \\

Equation (\ref{eq:ton_window}) explicitly takes \textit{b} into account, differently from Eq. (14) in \cite{parvez2010analytic}, where $b=1$ is assumed. 
We will show in Section \ref{sec:ton_performance} that, because of the finer estimations of $S^{FR}$ and $\mathbb{E}[\delta]$, the expected value of CWND in (\ref{eq:ton_window}) is closer to the simulated one than the other theoretical models under consideration \cite{padhye2000modeling, parvez2010analytic, dunaytsev2007analytical}.

Each RTT is supposed as a r. v., whose value does not depend on CWND, as in \cite{padhye2000modeling, parvez2010analytic, dunaytsev2007analytical}.
We recall that: 
\begin{equation}
  D^{CA} = \mathbb{E}[X]+\mathbb{E}[D^\beta],
  \label{eq:ton_dCA}
\end{equation}
where $D^{CA}$ is the average duration, in number of rounds, of the CA phase, including the time spent waiting for entering in FR. Moreover, $D^{FR}$ is the average duration, in number of rounds, of the FR phase:
\begin{equation}
  D^{FR}=\mathbb{E}[\delta].
  \label{eq:ton_dFR}
\end{equation}
In fact, NewReno recovers one lost segment per round. Finally, $\mathbb{E}[D^{\beta}]$ is the expected time, in number of rounds, after the first loss and before entering FR; it can be approximated as $\mathbb{E}[D^{\beta}] \approx 1/2$.
From (\ref{eq:ton_dCA}) and (\ref{eq:ton_dFR}), the expected duration of a cycle $D^{CAFR}=\mathbb{E}[D^{CAFR}_i]$ [s], results as:
\begin{equation}
  D^{CAFR} = \mathbb{E}[RTT]\big(D^{CA}+D^{FR}\big).
  \label{eq:ton_time}
\end{equation}

\begin{figure}
  	\centering
  	\includegraphics[scale=0.37, clip=true, trim=0 0 0 0]{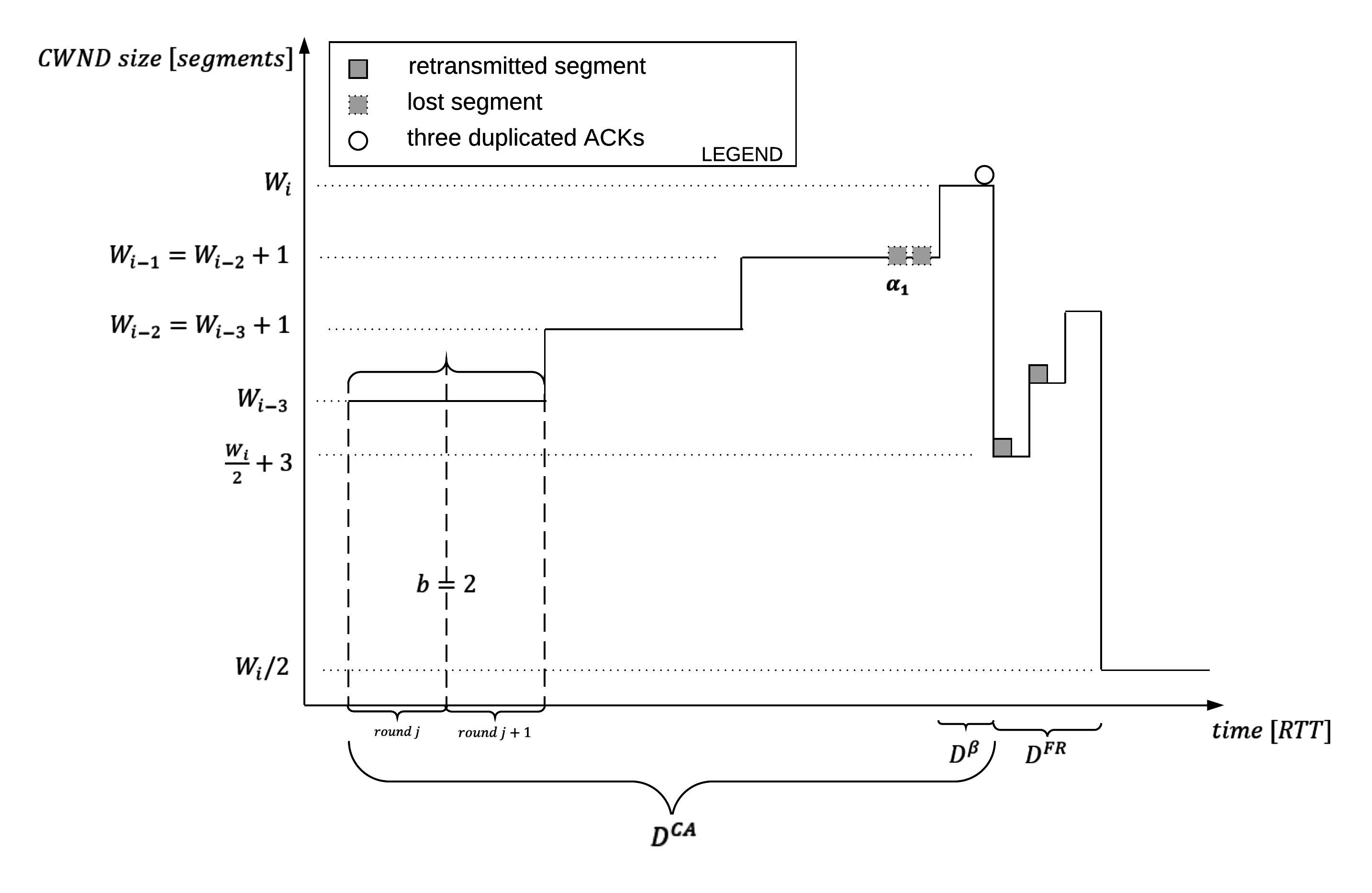}
	\caption{Evolution of CWND during CA and FR phases}
  	\label{fig:ton_wind2}
\end{figure}

Figure \ref{fig:ton_wind2} shows the evolution of CWND during CA and FR phases and the duration of the aforementioned phases. A CAFR cycle ends with the end of the FR phase because of one or more segment losses. The duration of the last part of the CA phase is $D^\beta$ RTTs: TD ACKs are received, prematurely ending the last round of the CA phase. The two segments lost in Figure \ref{fig:ton_wind2} are recovered during the FR phase immediately following, which lasts $D^{FR}=2$ RTTs. The CWND size is $(W_i/2 + 3)$ at the start of the FR phase and $W_i/2$ at the end of the FR phase.

Combining (\ref{eq:ton_segmentsent}) and (\ref{eq:ton_time}), we can finally write the throughput expression [segments/s], neglecting TOs, as:
\begin{equation}
   T_{noTO}=\frac{(\mathbb{E}[W]-4) \thinspace q+\frac{1}{p}}{\mathbb{E}[RTT]\left\{ b (\frac{\mathbb{E}[W]}{2} +1) + \frac{1}{2} + [1+(\mathbb{E}[W]-4) \thinspace q] \right\}}
	\label{eq:ton_Tnoto}
\end{equation}
where $\mathbb{E}[W]$ is expressed in (\ref{eq:ton_window}).

\subsection{Full Model}
\label{subsec:ton_Full}
This section extends the model in Section \ref{subsec:ton_noTO} to include also TOs as loss indications. Each cycle consists of either a CA or a CAFR period followed by a TO with an SS period. TCP NewReno may experience a TO either during CA or during FR, given that a loss event $(L)$ occurred with probability $p$. The former transition occurs with probability $p_{TOCA}=P\{TO\:|\:CA,\:L\}$, when TCP does not receive enough duplicate ACKs to trigger FR, while the latter transition occurs with probability $p_{TOFR}=P\{TO\:|\:FR,\:L\}$, when retransmitted segments are lost during the FR phase.
TCP can experience a TO in the CA phase when more than $(W-3)$ segments are lost in a drop window. 
The conditional probability $p_{TOCA}$ is given by:
\begin{equation}  
  p_{TOCA} = \mathbb{E} \left[ \sum\limits_{j=W_i-2}^{W_i} \binom{W_i - 1}{j - 1} (1-q)^{W_i-j}q^{j-1} \right].
  \label{eq:ton_ptoca}
\end{equation}

A TO occurs during FR if any of the retransmitted segments is lost. 
This condition can be approximated by assuming that if a loss event occurs in FR, then also the retransmitted segment is lost, thus triggering a TO \cite{parvez2010analytic}. For $\delta$ losses in the drop window, NewReno takes $\delta$ RTTs to recover the lost segments, during which it sends $S^{FR}$ segments. The probability that the $i$-th segment is lost, given that the previous $(i-1)$ segments are successfully delivered, is $(1-p)^{i-1}p$. Therefore, according to \cite{parvez2010analytic}, it follows that:
\begin{equation} 
  p_{TOFR} = \mathbb{E} \left[ \sum\limits_{j=1}^{W_i-3} B[W_i - 3, j] \left[ 1-(1-p)^{j \frac{W_i}{2}} \right] \right].
  \label{eq:ton_ptofr}
\end{equation}

During a TO, TCP does not send any segments. According to \cite{padhye2000modeling}, the duration of a TO period is given by:
\begin{equation}  
	D^{TO} = RTO\frac{1+\sum\limits_{j=0}^5 2^j p^{j+1}}{1-p},
	\label{eq:ton_dto}
\end{equation}
where the initial value of the $RTO$ parameter of TCP is 2 or 3 [s], in recent implementations\footnote{As in RFC 6298, \textit{Appendix A}.}.

In the SS phase, the initial CWND size is one and it grows until the SST, $W/2$, is reached. Therefore, in the last round of SS, TCP transmits $\mathbb{E}[W]/2$ segments, on average, then enters the CA phase. The rate of increase of the CWND, when the \textit{delayed ACKs} feature is in use, is $x=(1+\frac{1}{b})$ on an RTT basis \cite{cardwell2000modeling}:
\begin{equation} 
  \begin{split}
      S^{SS}&=\mathbb{E}\left[x^0 + x^1 + ... + x^{\log_{x} \frac{W_i}{4}}\right]=\mathbb{E}\left[ b \thinspace \left( \thinspace x \thinspace \frac{W_i}{4} - 1 \thinspace \right)\right] \\
      & = b \thinspace \left( \thinspace x \frac{\mathbb{E}[W]}{4} - 1 \right).
      \label{eq:ton_SS}
    \end{split}
\end{equation}
$S^{SS}$ represents the expected value of the number of segments sent during the SS phase as in \cite{parvez2010analytic}, where we consider the segments of the last round of the SS phase as being part of the CA phase. 
$D^{SS}$ represents the duration of the SS phase as proposed in \cite{parvez2010analytic}, but here explicitly considering $b$:
\begin{equation}
  \begin{split}
	D^{SS} &= \mathbb{E}\left[ RTT_i \left( \log_{x} \frac{W_i}{4}+1 \right) \right]\\
    & \approx \mathbb{E}[RTT] \left( \log_{x} \frac{\mathbb{E}[W]}{4}+1 \right).
    \end{split}
    \label{eq:ton_dSS}
\end{equation}
Note that the expected value in (\ref{eq:ton_dSS}) is applied to a non-linear expression. We have already assumed $RTT_i$ and $W_i$ as independent r. v.. Moreover, if the distribution of $W$ is concentrated around its mean, as shown in Section \ref{sec:ton_performance}, then the approximation on the derivation of the expected value in (\ref{eq:ton_dSS}) is acceptable.

The final expression for the throughput estimation, when considering also TOs, is given in (\ref{eq:ton_Tfull}). It is worth remarking that some parameters, for instance $\mathbb{E}[W]$ provided in Section \ref{subsec:ton_noTO}, are used also here, but they only consider CA and FR phases, without taking SS into account, thus providing an approximated estimate.
The formulation in (\ref{eq:ton_Tfull}) is presented in its whole expression, without any approximations. We will show in Section \ref{sec:ton_performance} that (\ref{eq:ton_Tnoto}) represents a throughput estimation that shows a very small error, so that the improvement provided by (\ref{eq:ton_Tfull}) is negligible.
\begin{table*}
  \begin{equation}
    T_{Full}=\frac{(1-p_{TOFR}-p_{TOCA})(S^{CA}+S^{FR})+p_{TOCA}(S^{SS}+S^{CA})+p_{TOFR}(S^{SS}+S^{CA}+S^{FR})}{(1-p_{TOFR}-p_{TOCA})(D^{CA}+D^{FR})+p_{TOCA}(D^{SS}+D^{CA}+D^{TO})+p_{TOFR}(D^{SS}+D^{CA}+D^{FR}+D^{TO})}.
    \label{eq:ton_Tfull}
  \end{equation}
\end{table*}
\section{The Relation between Burst Loss Rate and Event Loss Rate: the BLR Model}
\label{sec:ton_pl}

In this Section, a simple but effective model is proposed in order to glue together the Burst Loss Rate (BLR) value experienced at MAC layer and the $p$ and $q$ values at transport layer. BLR is the rate of losses due to collisions. In \cite{crdsa}, an analytical recursive model is proposed to derive an upper bound for the throughput of 2-CRDSA, but, to the best of the authors' knowledge, no analytical model is available for estimating throughput and BLR of CRDSA++.

In the following, we use the term \textit{collision} to refer to MAC bursts marked as lost after SIC, thus not recoverable in any way at the receiver. TCP segments are retransmitted if they are lost because of collisions. For the sake of simplicity, in the following we assume that one segment fits exactly into one time-slot (case $r=1$). We recall that buffer overflow effects are negligible because of the large buffer in use.
In the simulation results presented in Section \ref{sec:ton_performance}, each TCP retransmission corresponds to a collision in the channel. In our study and simulations, we neglect spurious TCP retransmissions. 

In what follows, a very simple expression is provided, denoted as the \textit{BLR model}, which links together $q$, $p$ and $BLR$. This expression holds if no spurious retransmissions occur and $f=1$. We recall that $p$ is the rate of loss events, $q$ corresponds to SLR and $\mathbb{E}[\delta]$ is the expected number of segments lost in a loss event. Each collision is responsible of a single segment loss, triggering a retransmission at transport layer. Thus, we can write:
\begin{equation}
  q \equiv BLR
  \label{eq:ton_blr_q}
\end{equation}
and, for a finite CWND size:
\begin{equation}
  p = \frac{q}{\mathbb{E}[\delta]} \approx \frac{q}{1+(\mathbb{E}[W]-4) \thinspace q} \approx \frac{BLR}{1+(\mathbb{E}[W]-4) \thinspace BLR}.
  \label{eq:ton_blr_p}
\end{equation}
Plugging (\ref{eq:ton_blr_q}) and (\ref{eq:ton_blr_p}) in (\ref{eq:ton_Tnoto}) and (\ref{eq:ton_Tfull}), we obtain the throughput estimation as a function of BLR. Following this approach, it is sufficient to know the collision rate at MAC layer, i.e., BLR, to obtain $q$ and $p$ rates to derive a TCP throughput estimation. In Section \ref{sec:ton_blrmodel}, the \textit{BLR model} is validated against simulation results.
\section{Performance Evaluation}
\label{sec:ton_performance}

The performance evaluation shown here is based on the NS-3 simulator, implementing a DVB-S2/RCS2 satellite network by means of the modules described in \cite{sns3}. The TCP flavor in use is NewReno. The available implementation in NS-3 is based on the SBS variant. Note that, as shown in \cite{dunaytsev2007analytical}, the performance of SBS is not significantly different from that of the Impatient variant in the conditions here under consideration, i.e., limited congestion window sizes and independent loss events. The main settings of the simulator are provided in Table \ref{table:ton_settings}.
\begin{table}
\caption{System settings}
  \begin{center}
  \begin{tabular}{|c|c|} \hline
   \textbf{Name} & \textbf{Value} 			\\ \hline \hline   
  TCP flavor & NewReno						\\ \hline
  TCP MSSs & 23 / 173 [B]					\\ \hline
  TCP/IP headers size & 7 [B] (w/ ROHC) 	\\ \hline
  Initial RTO value & 2 [s]					\\ \hline
  RA scheme & CRDSA++ (3 replicas)			\\ \hline 
  RA blocks per superframe & 1				\\ \hline
  RA block duration & 13 [ms]				\\ \hline
  Time-slots per RA block, WF 14 & 64		\\ \hline
  Time-slots per RA block, WF 3 & 194		\\ \hline
  Bandwidth & 8012820 [Hz]					\\ \hline
  Roll off & 0.2							\\ \hline
  Carrier spacing & 0.3 [Hz]				\\ \hline
  Nominal Round Trip Time & 0.52 [s]		\\ \hline  
  \end{tabular}
  \end{center}
  \label{table:ton_settings}
\end{table}
The results here intend to assess how changing WF and MSS values can alter the overall performance. The main aim of this study is in clarifying if TCP is a suitable choice as transport protocol over a long delay RA link and, if this is the case, which choice of parameters leads to the highest throughput.

We recall that the $G^*$ load, i.e., the normalized MAC load level achieving the highest throughput at MAC level, depends on the number of available time-slots per RA block \cite{bacco2013generalized}, among other factors. Thus, while keeping the other parameters unmodified, the choice of a different waveform can change the number of available time-slots per RA block, shifting $G^*$. For instance, when 64 time-slots are available, the $G^*$ value of CRDSA++ is $\approx 0.7$ [bursts/time-slot]; when 194 time-slots are available, the $G^*$ value of CRDSA++ is $\approx 0.78$ [bursts/time-slot] \cite{bacco2013generalized}.
The overall system has been tested at several increasing load levels, before, close to and after $G^*$. Two crucial aspects are characterized in this work: the MAC operating point $\hat{G}_T$ when NewReno is in use over an RA satellite channel, and the system stability; both issues are discussed in Section \ref{sec:ton_reg}. To the best of the authors' knowledge, the characterization of $\hat{G}_T$ is not available in literature for CRDSA++, if we exclude our preliminary contribution in \cite{bacco2015tcp}.

The focus in what follows is on normalized load levels $G \geq 0.45$, because, up to this point, CRDSA++ throughput is almost equal to the offered load (i.e., the operating point is in the linear part) \cite{del2009high}; beyond this load level, the system works in the quasi-linear part up to the $G^*$ point, and then the MAC throughput dramatically decreases because of a too large BLR. If a DVB-RCS2 satellite system with ARQ enabled (typical setting) works at $G > G^*$ and no load control algorithm is used, the system may experience instability \cite{meloni2014random}.

\subsection{Throughput validation}
\label{sec:ton_estth}
The NewRenoSAT model described in Section \ref{sec:ton_model} is here validated against simulation results and compared with other approaches in literature (see Section \ref{subsec:ton_related}): (\textit{i}) the contribution in \cite{padhye2000modeling}, then amended in \cite{towcomments} (here denoted as PFTK model); (\textit{ii}) the contribution in \cite{parvez2010analytic} (here denoted as PWM model); (\textit{iii}) the contribution in \cite{dunaytsev2007analytical} (here denoted as DAKH model). 
In the following, the provided throughput estimations are obtained by using simulated $p$ and $q$ values because there is no analytical model in literature able to characterize the layer-2 losses when CRDSA++ is in use.

\subsubsection{Accuracy of the CWND estimation}
\label{sec:ton_cwnd}
\begin{figure*}
    \centering
        \begin{subfigure}[t]{0.3\textwidth}
          \includegraphics[scale=0.37, clip=true, trim=70 0 0 0]{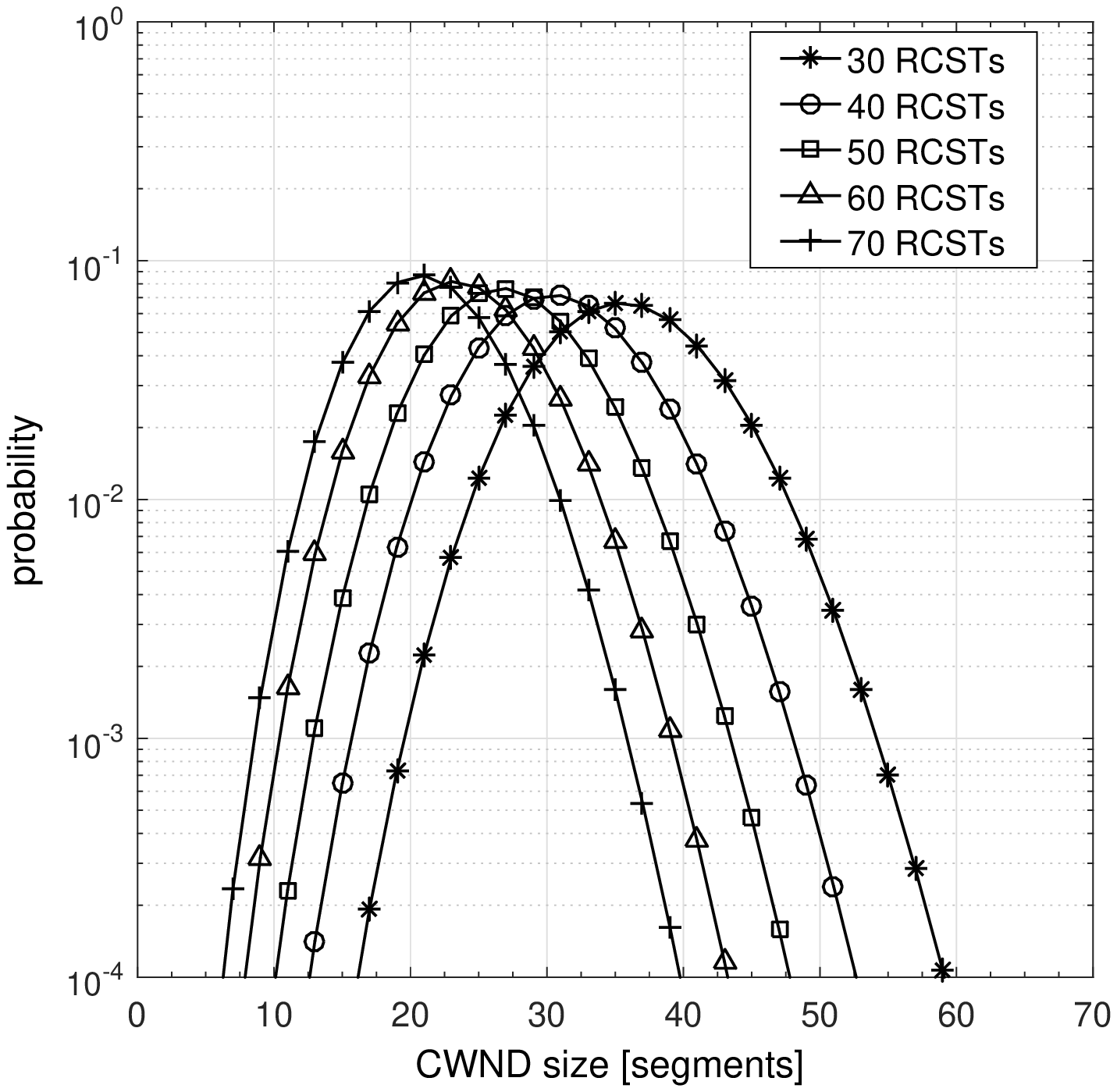}
          \caption{Probability Mass Function of the CWND size}
          \label{fig:ton_cwnddist}
    	\end{subfigure}
    ~
        \begin{subfigure}[t]{0.3\textwidth}
          \includegraphics[scale=0.238, clip=true, trim=25 5 0 45]{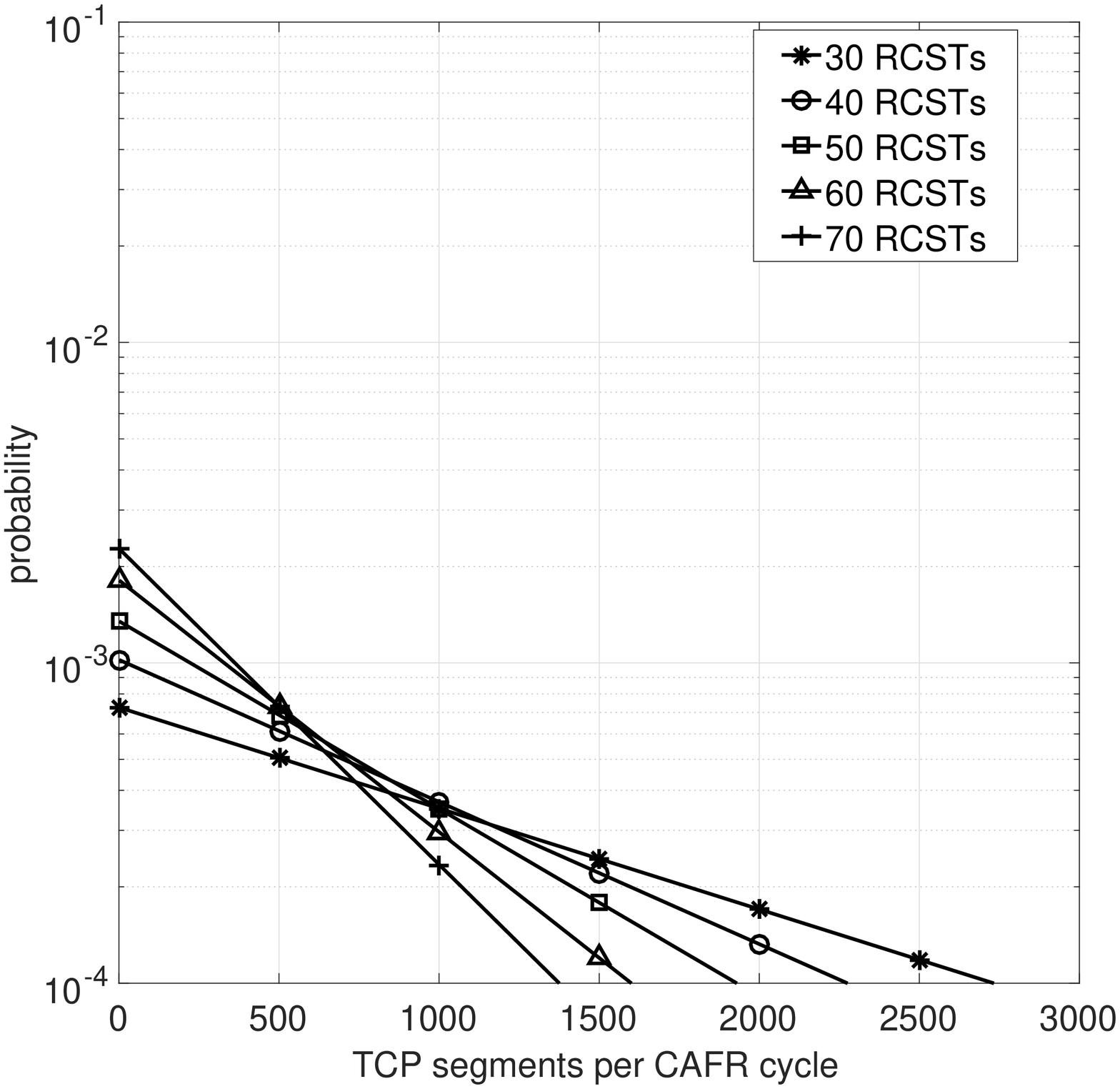}
          \caption{Probability Mass Function of the number of segments sent per CAFR cycle}
          \label{fig:ton_cycle}
    	\end{subfigure}
    ~ 
        \begin{subfigure}[t]{0.3\textwidth}
          \includegraphics[scale=0.37, clip=true, trim=70 0 0 0]{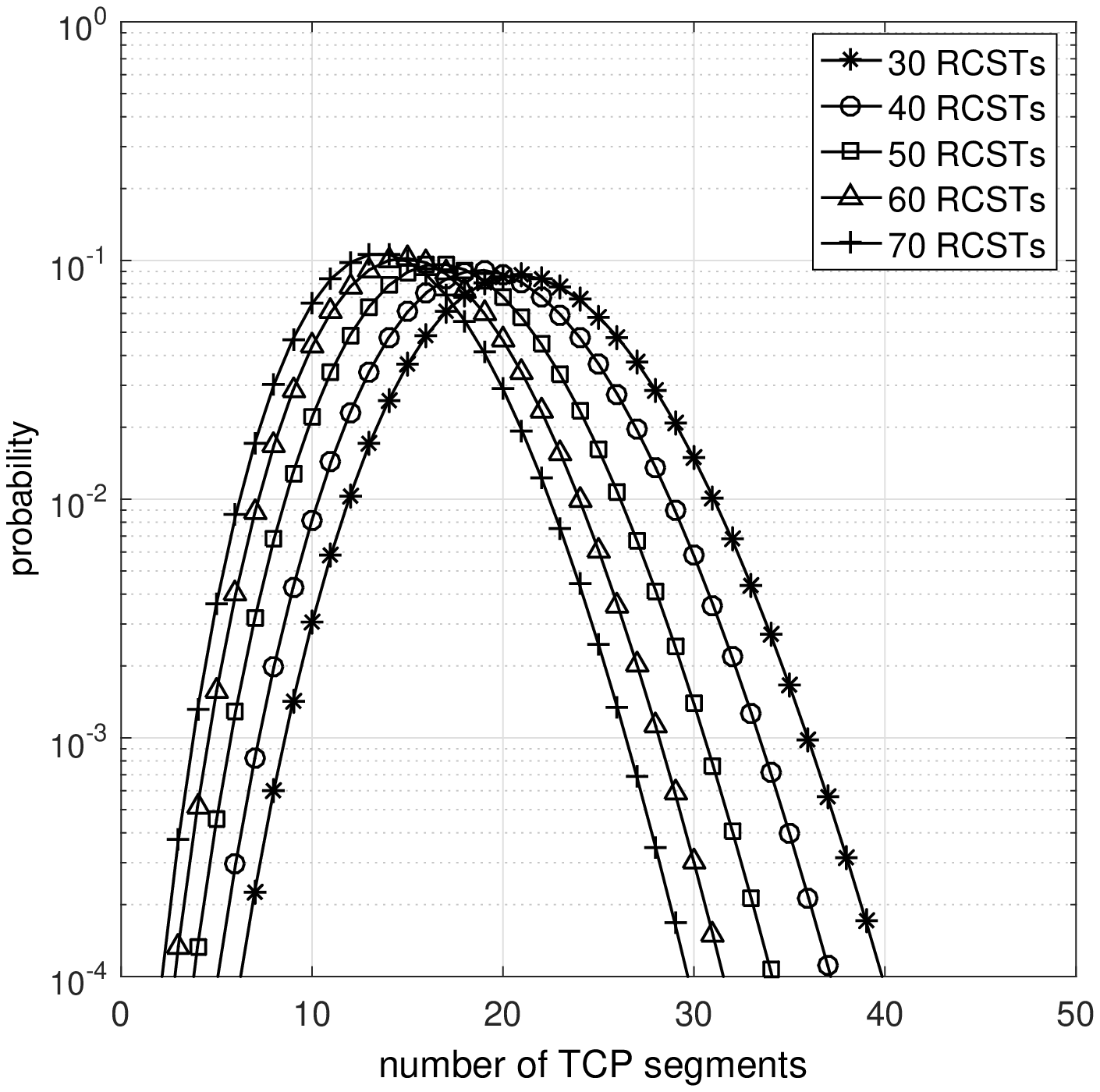}
          \caption{Probability Mass Function of the number of segments sent per FR phase}
          \label{fig:ton_frfr}
    	\end{subfigure}
    \caption{Simulation results of the scenario WF 14, MSS = 173 bytes, 64 time-slots per RA block}
    \label{fig:ton_14173_stats}
\end{figure*}

A critical value to be computed, when estimating the TCP throughput, is the CWND size. While just the expected value is used for the throughput estimation, the CWND exhibits a range of different values during the whole TCP connection lifetime. Figure \ref{fig:ton_cwnddist} shows the distribution of CWND obtained from simulations, when WF 14 and MSS = 173 bytes are selected; in the following, this scenario is used as exemplary case. 
As the load increases (i.e., when the number of RCSTs increases), $\mathbb{E}[W]$ decreases, while the distribution keeps the same shape. A larger $G$ value means higher load on the RA channel and then higher BLRs. Each time two or more RCSTs are involved in a collision, their CWND size is shrunk to a lower value, as in Figure \ref{fig:ton_wind2} after the two lost segments. Therefore, $\mathbb{E}[W]$ decreases as $G$ increases.
In Table \ref{table:ton_newrenosatVSsims}, the simulation results are compared with the NewRenoSAT model estimations, showing a close match.
\begin{table*}
	\caption{NewRenoSAT estimated values vs. simulation results\\WF 14, MSS = 173 bytes, 64 time-slots per RA block}
    \begin{center}
    \begin{tabular}{|c|c|c|c|c|c|c|} \hline
	\#RCSTs 
	&\multicolumn{2}{c|}{$\mathbb{E}[W]$} 
    &\multicolumn{2}{c|}{$S^{CA}$ [segments], $D^{CA}$ [s]} 
    &\multicolumn{2}{c|}{$S^{FR}$ [segments], $D^{FR}$ [s]}
    \\ \cline{2-3} \cline{4-5} \cline{6-7}    
    & simulated & model & simulated & model & simulated & model 
    \\ \hline    
	30 & 40.15 & 41.8 & 1367.6, \thinspace 25.5 & 1394, \thinspace 28.4 & 19.3, \thinspace 0.62 & 20.5, \thinspace 0.66 \\ \hline
    40 & 33.4 & 34.8 & 1003.7, \thinspace 21.4 & 978, \thinspace 22.4 & 15.6, \thinspace 0.58 & 17.1, \thinspace 0.63 \\ \hline
    50 & 28.5 & 29.6 & 721.4, \thinspace 17.9 & 716.6, \thinspace 18.6 & 13.2, \thinspace 0.58 & 14.4, \thinspace 0.61 \\ \hline
    60 & 26.1 & 26 & 552.1, \thinspace 15.6 & 559.2, \thinspace 16.5 & 11.7, \thinspace 0.58 & 12.6, \thinspace 0.61 \\ \hline
    70 & 21.6 & 21.9 & 433.3, \thinspace 13.9 & 406.1, \thinspace 14.2 & 10.4, \thinspace 0.58 & 10.6, \thinspace 0.62 \\ \hline    
    \end{tabular}
    \end{center}
  \label{table:ton_newrenosatVSsims}
\end{table*}

The average value of the CWND has a direct impact on the number of segments sent in a CAFR phase, as visible in (\ref{eq:ton_esi}). The distribution of the number of segment sent per CAFR phase is shown in Figure \ref{fig:ton_cycle}: this is a geometric distribution, as pointed out in \cite{padhye2000modeling}. In a CAFR cycle, the sending rate varies if the connection is in the CA phase or in the FR phase: to underline this, Figure \ref{fig:ton_frfr} shows the distribution of the number of segments sent in FR only. The distribution in the FR phase (in Fig. \ref{fig:ton_frfr}) is different from the one in the CAFR phase (in Fig. \ref{fig:ton_cycle}): the CA phase completely dominates the dynamics of a CAFR cycle, and it exhibits an approximate geometric distribution, while the FR phase has quite a negligible impact on the overall distribution, thus the approximation of a geometric distribution still stands. 
Table \ref{table:ton_newrenosatVSsims} also shows how the NewRenoSAT model is able to accurately estimate the average number of segments in CA and FR phases, along with their average durations.

\subsubsection{Average number of segment losses per loss event}
\label{sec:ton_delta}
A central part of this work is the characterization of the loss process experienced on the RA channel when CRDSA++ is in use. Equation (\ref{eq:ton_delta}) provides the analytical expression of $\mathbb{E}[\delta]$, the expected number of losses per loss event. It ranges from 1 to 1.2 in our simulation results, as shown in Table \ref{table:ton_comparison_final}, if we exclude the case of WF 14 and MSS = 23 bytes, later discussed. It means that, on average, 1 to 1.2 segments are lost per drop window: the same number of segments is retransmitted in the FR phase, which follows the loss event. Because of the reduced available bandwidth per RCST, the average CWND size is small and, on average, a single loss is experienced per drop window if $q$ is small. 

Instead, a strong correlation effect is present in the scenario WF 14 and MSS = 23 bytes: a single time-slot carries up to $r=5.7$ segments, thus a single collision causes a burst of $f=6$ losses at transport layer. Note that $r$ is the maximum number of whole or partial TCP segments into a single time-slot, and not the actual one, because the RLE protocol always tries to ensure the largest occupation of the time-slot, but it is limited by the number of data units actually pending for transmission. It means that, if less than $f$ data units are in the queue, a smaller number will be transmitted in a single time-slot, thus sub-utilizing the available resources.
Moving to the performance evaluation, in this case a single collision represents a very stressful event for a connection, because of the burst of losses experienced at TCP layer. This scenario exhibits very low performance because of this effect, when compared with the scenario WF 14 and MSS = 173 bytes and, generally, its performance are worse than any other configuration with $f=1$. Because of this, we are not intended to the study of the TCP behavior under the hypothesis of bursty losses due to encapsulation effects (i.e., $f>1$). For the same reason, Table \ref{table:ton_comparison_final} does not show the $\mathbb{E}[\delta]$ value, whose analytical expression is valid only under the hypothesis of independent losses.

\subsubsection{Comparison among different approaches for estimating the TCP throughput}
\label{sec:ton_thrp_est}
\begin{figure}[!ht]
    \centering
    \includegraphics[scale=0.36, clip=true, trim=0 50 0 50]{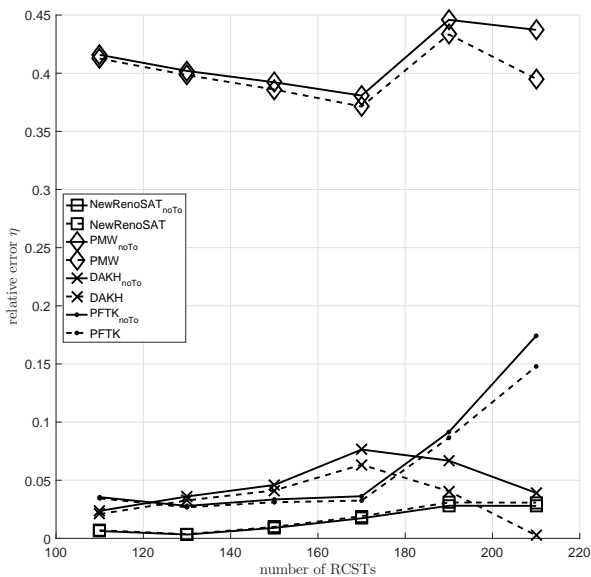} 
    \caption{Comparison (relative error) between simulations and theoretical approaches for TCP NewReno throughput over CRDSA++ for WF 3, MSS = 23 bytes and 194 time-slots per RA block}
    \label{fig:ton_th3_23}
\end{figure}

\begin{figure}[!ht]
    \centering
    \includegraphics[scale=0.36, clip=true, trim=0 50 0 50]{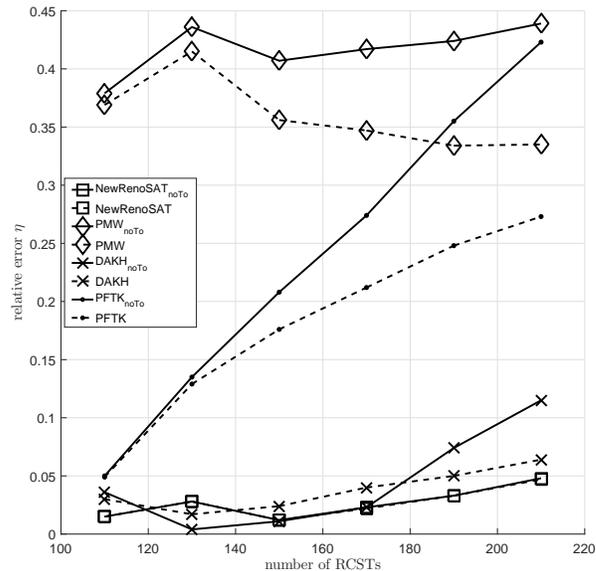}
    \caption{Comparison (relative error) between simulations and theoretical approaches for TCP NewReno throughput over CRDSA++ for WF 3, MSS = 173 bytes and 194 time-slots per RA block}
    \label{fig:ton_th3_173}
\end{figure}

\begin{figure}[!ht]
    \centering
     \includegraphics[scale=0.4, clip=true, trim=0 70 0 50]{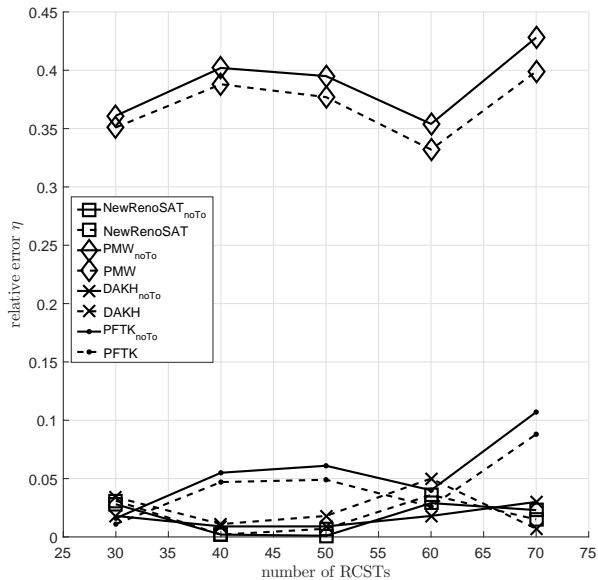}
    \caption{Comparison (relative error) between simulations and theoretical approaches for TCP NewReno throughput over CRDSA++ for WF14, MSS = 173 bytes and 64 time-slots per RA block}
    \label{fig:ton_th14_173}
\end{figure}

In Figures \ref{fig:ton_th3_23}, \ref{fig:ton_th3_173} and \ref{fig:ton_th14_173}, we compare the aforementioned TCP models for the analysis of TCP throughput over a RA satellite link using the CRDSA++ protocol. The throughput estimation relies on the use of $p$, $q$ and $\mathbb{E}[RTT]$ values from simulations, as in Table \ref{table:ton_comparison_final}.

The relative error $\eta= \lvert 1-\dfrac{T_{est}}{T_{sim}} \rvert$ of the estimations is shown, where $T_{est}$ is the throughput resulting by the analytical models under consideration \cite{padhye2000modeling, parvez2010analytic, dunaytsev2007analytical} and NewRenoSAT, and $T_{sim}$ is the throughput resulting from the simulations. The results here shown are obtained with extensive Montecarlo simulation runs: more than 1 million sent segments per scenario under consideration, with a narrow confidence interval.
$\eta$ is computed for every combination of WF and MSS in use, except for the scenario WF 14 and MSS = 23 bytes ($f > 1$), as explained before. Let us recall that $b=2$ (delayed ACKs) is used in the simulations in this work; the PMW model does not account for it, so the relative error of this analytical model is larger when compared with the others under consideration.

Figure \ref{fig:ton_th3_23} shows the value of $\eta$ for the scenario WF 3 and MSS = 23 bytes: 194 time-slots are available for transmission and an RCST can use one time-slot per RA block. A segment fits exactly into one MAC burst ($r=1$). The NewRenoSAT model proposed in this work show the lowest $\eta$ value (i.e., the best results), if we except the case with the largest number of RCSTs (i.e., the largest load); the DAKH model and the PFTK one are quite accurate, too, but the latter has a $\eta$ larger than 0.15 for higher loads and the former tends to overestimate the throughput at medium loads.

In Figure \ref{fig:ton_th3_173} the relative error for the scenario WF 3 and MSS = 173 bytes is shown. In this scenario, a time-slot carries a fraction of the segment: in fact, $r=0.175$, so that 5.7 time-slots are needed to carry a single segment, which means that the sixth time-slot carries the last fraction of a segment and the initial fraction of the next one. A collision involving the sixth time-slot causes the loss of two TCP segments, thus triggering two retransmissions for a single collision event. This happens with probability equal to $1/6$: because of this, the error on the estimation is larger than the one in Figures \ref{fig:ton_th3_23} and \ref{fig:ton_th14_173}, for each model. In this scenario, there is some correlation among losses, so the hypothesis of complete independence among segment losses is not verified. Anyway, the correlation is limited, so the simulation results show a good match with NewRenoSAT and DAKH analytical models anyway.

Figure \ref{fig:ton_th14_173} provides the throughput estimation for the scenario WF 14 and MSS = 173 bytes: one segment fits exactly into a single MAC burst ($r=1$). NewRenoSAT and DAKH models achieve the best results, even if DAKH shows an average relative error that is larger at high loads. The PFTK model exhibits a larger error, instead; in fact, it does not take the FR mechanism into account.

\subsubsection{Retransmission TimeOuts}
In this Section, the measure $\xi$ is introduced to quantify the rate of TOs in the simulations: 
\begin{equation}
	\xi = \frac{ \mathbb{E}[\#TOs] }{ T_s / \mathbb{E}[RTT] }.
	\label{eq:ton_xi}
\end{equation}
$\mathbb{E}[\#TOs]$ is the average number of TOs per connection and $T_s$ is the simulation length in seconds, where $T_s / \mathbb{E}[RTT]$ is the number of rounds in each simulation. Therefore, $\xi$ represents the probability that a TO occurs in a round. The numerical values of $\xi$ per scenario under consideration are shown in the last column of Table \ref{table:ton_comparison_final}.

We observe two effects worth of mentioning: if $r=1$, $\xi$ has a substantially lower value with respect to the case where $r \neq 1$; then, MAC fragmentation and packing significantly impact on TCP performance. The second effect here discussed is related to the load level: the lowest simulated load level exhibits a $\xi$ value larger than that at higher ones, which is unexpected, at a first glance.  At high loads, the collision rate is responsible of sustained segment losses, thus the $\xi$ value reflects this. 
At medium loads, the collision rate is responsible of moderate losses, thus $\xi$ shows a lower value. In fact, the FR mechanism is particularly efficient in presence of a moderate loss rate, and time-consuming TOs are avoided. Instead, at low loads, another effect is present, which makes $\xi$ larger than expected: the CWND overshooting problem. A collision is a very rare event at low loads, therefore the overshooting problem is possible: a CWND that increases slowly but continuously over time and whose value goes over the Bandwidth-Delay product increases the probability of spurious TOs.

\begin{table*}
 \caption{Simulation results: the scenarios are described in the first column.\\$\mathbb{E}[RTT]$ is measured in seconds, $Thr.$ is the average simulated throughput [kbps] per TCP connection}
 \begin{center}
 \begin{tabular}{|c|c|c|c|c|c|c|c|c|c|c|} \hline
 
\textbf{$WF, \thinspace MSS$} & \textbf{$\#RCSTs$} & \textbf{$BLR$} & \textbf{$r$} & \textbf{$f$} & \textbf{$q$} & \textbf{$p$} & \textbf{$\mathbb{E}[\delta]$} & \textbf{$\mathbb{E}[RTT]$} & \textbf{$Thr.$} & \textbf{$\xi$} \\ \hline \hline

	3, \thinspace 23 & 110 & 1.1E-4 & 1 & 1 & 1.44E-3 & 4.5E-4 & 1.07 & 0.69 & 12.50 & 5.5E-4 
    \\ \hline 
    3, \thinspace 23 & 130 & 2E-4 & 1 & 1 & 1.70E-3 & 4.86E-4 & 1.08 & 0.68 & 11.90 & 3E-5 
    \\ \hline
    3, \thinspace 23 & 150 & 5.5E-4 & 1 & 1 & 2.80E-3 & 6.68E-4 & 1.11 & 0.62 & 9.60 & 6.2E-5 
    \\ \hline 
    3, \thinspace 23 & 170 & 9.2E-4 & 1 & 1 & 5.13E-3 & 9.86E-4 & 1.16 & 0.61 & 7.84 & 5E-4 
    \\ \hline 
    3, \thinspace 23 & 190 & 1.1E-3 & 1 & 1 & 7.94E-3 & 1.20E-3 & 1.20 & 0.60 & 6.86 & 2.5E-3 
    \\ \hline 
    3, \thinspace 23 & 210 & 2.9E-3 & 1 & 1 & 1.1E-2 & 6.3E-3 & 1.20 & 0.59 & 2.90 & 4E-3 
    \\ \hline 
	\hline
    3, \thinspace 173 & 110 & 1.2E-4 & 0.175 & 1 & 4.1E-3 & 1.5E-3 & 1.09 & 2.40 & 11.94 & 1E-1 
    \\ \hline 
    3, \thinspace 173 & 130 & 3.5E-4 & 0.175 & 1 & 8.66E-3 & 3.5E-3 & 1.14 & 1.50 & 11.90 & 9E-2 
    \\ \hline 
    3, \thinspace 173 & 150 & 1.4E-3 & 0.175 & 1 & 2.35E-2 & 1.1E-2 & 1.16 & 0.92 & 10 & 7E-2 
    \\ \hline 
    3, \thinspace 173 & 170 & 2.8E-3 & 0.175 & 1 & 2.94E-2 & 1.72E-2 & 1.17 & 0.77 & 8.50 & 5E-2 
    \\ \hline 
    3, \thinspace 173 & 190 & 4.4E-3 & 0.175 & 1 & 3.3E-2 & 2.7E-2 & 1.18 & 0.72 & 7.30 & 6E-2 
    \\ \hline 
    3, \thinspace 173 & 210 & 6.16E-3 & 0.175 & 1 & 3.5E-2 & 3.5E-2 & 1.20 & 0.69 & 6.37 & 7E-2 
    \\ \hline \hline
    
    14, \thinspace 23 & 30 & 2.27E-3 & 5.7 & 6 & 1.3E-2 & 3.7E-4 & - & 1.28 & 10.5 & 5E-2 
    \\ \hline 
    14, \thinspace 23 & 40 & 2E-3 & 5.7 & 6 & 1.38E-2 & 3.8E-4 & - & 1.275 & 10.48 & 4E-2 
    \\ \hline 
    14, \thinspace 23 & 50 & 2.32E-3 & 5.7 & 6 & 1.64E-2 & 4.2E-4 & - & 1.23 & 10.1 & 4E-2 
    \\ \hline 
    14, \thinspace 23 & 60 & 2E-3 & 5.7 & 6 & 1.71E-2 & 4.3E-4 & - & 1.19 & 9.93 & 4E-2 
    \\ \hline 
    14, \thinspace 23 & 70 & 1.9E-3 & 5.7 & 6 & 1.76E-2 & 4.7E-4 & - & 1.15 & 9.72 & 4E-2 
    \\ \hline \hline 
    
    14, \thinspace 173 & 30 & 6.41E-4 & 1 & 1 & 8.52E-4 & 7.45E-4 & 1.03 & 0.64 & 65.75 & 5.3E-4 
    \\ \hline
    14, \thinspace 173 & 40 & 9.26E-4 & 1 & 1 & 1.34E-3 & 1E-3 & 1.04 & 0.60 & 58.50 & 1E-4 
    \\ \hline
    14, \thinspace 173 & 50 & 1.31E-3 & 1 & 1 & 1.75E-3 & 1.37E-3 & 1.04 & 0.58 & 51.51 & 1.4E-4 
    \\ \hline 
    14, \thinspace 173 & 60 & 1.76E-3 & 1 & 1 & 2.30E-3 & 1.79E-3 & 1.05 & 0.58 & 45.20 & 3E-4 
    \\ \hline 
    14, \thinspace 173 & 70 & 2.3E-3 & 1 & 1 & 3.5E-3 & 2.28E-3 & 1.07 & 0.58 & 38.11 & 4.6E-4 
    \\ \hline 
 
 \end{tabular}
 \end{center} 
 \label{table:ton_comparison_final}
\end{table*}

\subsection{Validating the BLR model}
\label{sec:ton_blrmodel}
The \textit{BLR model} discussed in Section \ref{sec:ton_pl} is now validated against simulation results. The throughput estimation here presented relies on the use of $BLR$ and $\mathbb{E}[RTT]$ values taken from simulations, as in Table \ref{table:ton_comparison_final}. The scenario WF 14 and MSS = 173 bytes has been chosen because of the very low rate of spurious retransmissions. 

Table \ref{table:ton_BLRmodelvalidation} shows the relative error $\eta$ when comparing the simulation results with the estimations provided by (\ref{eq:ton_Tnoto}) and (\ref{eq:ton_Tfull}) in the first two columns. Further to this, the accuracy of the \textit{BLR model} is visible in the last two columns, i.e., when plugging (\ref{eq:ton_blr_q}) and (\ref{eq:ton_blr_p}) in (\ref{eq:ton_Tnoto}) and (\ref{eq:ton_Tfull}). The \textit{BLR model} provides good precision: in fact, the larger estimation error is $\leq 7$\%. 
We recall that the \textit{BLR model} presented in Section \ref{sec:ton_pl} neglects spurious retransmissions, thus causing a small error in the throughput estimation.
A consideration is here in order to motivate the need of the \textit{BLR model}: while it can be practical to estimate BLR in real systems, the same cannot be said for $p$ and $q$ when using TCP. Thus, this model offers a simple and elegant way, in the authors' opinion, for estimating the TCP throughput over an RA satellite link.

\subsection{Stability}
\label{sec:ton_reg}
A critical issue to be taken into account, when dealing with RA protocols, is system stability. CRDSA++, as other RA protocols, has an optimal working point, namely $G^*$, which exhibits the maximum throughput offered by the MAC protocol, namely $T^*$. If the system is forced to work at loads $G > G^*$, instability may occur and proper countermeasures are needed, as analysed in \cite{meloni2014random}.

\begin{figure*}
    \centering
        \begin{subfigure}[t]{0.45\textwidth}
          \includegraphics[scale=0.4, clip=true, trim=35 60 0 80]{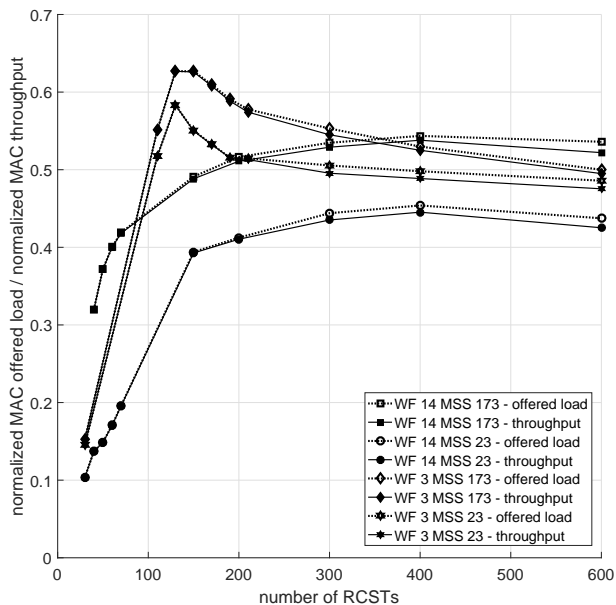}
          \caption{Aggregate normalized MAC throughput and aggregate offered load vs number of RCSTs}
          \label{fig:ton_autoreg_new}
    	\end{subfigure}
    ~
        \begin{subfigure}[t]{0.45\textwidth}
          \includegraphics[scale=0.4, clip=true, trim=45 80 0 80]{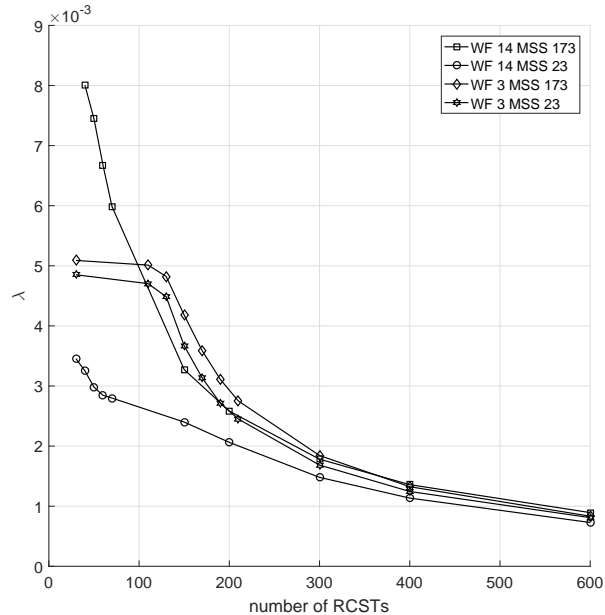}
          \caption{Offered load per RCST $\lambda$ vs number of RCSTs}
          \label{fig:ton_ptx}
    	\end{subfigure}
    \caption{MAC offered load and MAC throughput for different WFs and MSSs}
	\label{fig:ton_final}
\end{figure*}

DVB-RCS2 standard encompasses a normative load control algorithm, aimed at keeping the system at a target operating point $G_T$, where $G_T \leq G^*$. This load control algorithm is not used in our simulator because the work in \cite{Munari_IJSCN2014} shows that the aforementioned algorithm needs to be tuned in accordance with each scenario under consideration, and this can be not trivial at all; furthermore, a more complex algorithm would be necessary to target realistic Quality of Service (QoS) requirements.
On the contrary, the use of TCP ensures the stability at every load level under consideration, as discussed later in this part, and does not require any tuning. 
The DVB-RCS2 load control algorithm can limit the magnitude of the load oscillations around $G_T$ according to two different strategies \cite{Munari_IJSCN2014}: by reducing the number $N$ of RCSTs or by shifting $G_T$ to a lower value. These oscillations can be responsible for pushing the instantaneous offered load $G$ beyond $G^*$, thus leading the system towards the instability region; hence, controlling these oscillations is of primary importance. 
Both strategies offer disadvantages: in an IoT/M2M scenario, a large population of terminals (RCSTs) is common, so that reducing $N$ could be unfeasible. Instead, shifting $G_T$ towards a lower value can be a better solution in our scenario, but the use of a centralized NCC, which periodically monitors the aggregate load, is needed. The NCC is responsible for choosing the $G_T$ value according to the instantaneous $G$.
Figure \ref{fig:ton_autoreg_new}, later described, shows that the use of TCP removes the need of a centralized control: 
TCP can control $G_T$. 
Hence, TCP ensures the stability, without requiring any layer-2 control mechanisms that can improperly interact with the TCP congestion control algorithm. 

Figure \ref{fig:ton_autoreg_new} shows the number of RCSTs versus the normalized MAC throughput and normalized MAC offered load, when using TCP on top of WFs 14 and 3. We recall that WF 3 offers a greater number of time-slots carrying small payloads, while WF 14 offers a lower number of time-slots carrying larger payloads. Three load intervals can be read in Figure \ref{fig:ton_autoreg_new}, w.r.t. the number of RCSTs: the first one, at low loads, up to $\approx100$ RCSTs, where WF 14 offers a larger normalized throughput; the second one, at medium loads, from $\approx100$ to $\approx350$ RCSTs, where WF 3 outperforms WF 14 thanks to the greater number of time-slots; finally, the third interval, at larger loads, from $\approx350$ RCSTs on, where the throughput offered by the two WFs is almost comparable. 
A consideration is here in order: when using WF 3, TCP behaves in a greedy way, showing a clear load peak when $\approx150$ RCSTs are present, then the load decreases as the number of RCSTs increases. On the contrary, WF 14 shows a more balanced behavior, and a peak is not recognizable. The key difference is the number of available time-slots, which in turns determines the number of active RCSTs per RA block. The number of active RCSTs per RA block determines the SLR that each RCST experiences, because, for a given load $G$, the corresponding BLR can be estimated, and the BLR model in Section \ref{sec:ton_pl} can be used to relate BLR and SLR.

Let us call $\hat{G}_T$ the working point for TCP NewReno on top of CRDSA++ protocol. According to the y-axis values in Figure \ref{fig:ton_autoreg_new}, we can see that the following ranges are possible for the working point: (\textit{i}) $\hat{G}_T \in [0.45, \thinspace 0.55]$ for TCP over WF 14 (with 64 available time-slots); (\textit{ii}) $\hat{G}_T \in [0.58, \thinspace 0.63]$ for TCP over WF 3 (with 194 available time-slots).
When CRDSA++ (3 replicas) is in use and assuming no flow control algorithms, (i.e., TCP is not in use, as well), $G^*\approx0.7$, when 64 time-slots are available per RA block \cite{bacco2013generalized}, and $G^*\approx0.78$, when 194 time-slots are available per RA block \cite{bacco2013generalized}. 
Thus, as Figure \ref{fig:ton_autoreg_new} shows, $\hat{G}_T \leq G^*$: the segment loss rate and 
the congestion control algorithm shift $\hat{G}_T$ towards lower load levels, which exhibit lower loss rates.
This behavior is similar to the DVB-RCS2 normative load control, when $G_T$ is forced to a value lower than $G^*$: TCP congestion control algorithm leads $\hat{G}_T$ to oscillate around a stable equilibrium point without any centralized entity that adopts complex load control strategies.

We recall that $\lambda=G/N$ is the offered load per RCST.
Figure \ref{fig:ton_ptx} shows that, in the presence of few RCSTs (up to 100, approximately), a reduced number of larger time-slots (as for WF 14) should be preferred, in order to allow each RCST to take as much advantage as possible of the available resources. On the other hand, when the number of RCSTs increases, a WF that offers more time-slots (as for WF 3) should be preferred in order to allow a larger population to use the channel; in fact, even if $\lambda$ shows little value variations among 100 and 400 RCSTs, the aggregate throughput in Figure \ref{fig:ton_autoreg_new} for the same load range is substantially larger or, at least, equal. It is worth noting in Figure \ref{fig:ton_ptx} that $\lambda$ has approximately the same value for $N > 400$, independently from WF and MSS. 
Finally, Figure \ref{fig:ton_ptx} confirms that the scenario WF 14 and MSS = 23 bytes entails low performance and a large sub-utilization of available resources: thus, the choice of a WF and a MSS value which leads to $f>1$ should be avoided.
\begin{table*}
 	\caption{WF 14, MSS = 173 bytes, 64 time-slots per RA block. Relative error $\eta$ when comparing the accuracy of the model\\presented in Section \ref{sec:ton_model} and, in the last two columns, when plugging the \textit{BLR model} in (\ref{eq:ton_Tnoto}) and (\ref{eq:ton_Tfull})}
 	\centering
 	\begin{tabular}{|c|c|c|c|c|} \hline 
	\textbf{$\#RCSTs$} & \textbf{$NewRenoSAT_{noTO}$} & \textbf{$NewRenoSAT$} & \textbf{$BLR model_{noTO}$} & \textbf{$BLR model$} \\ \hline \hline
	30 & 0.028 & 0.031 & 0.063 & 0.069 \\ \hline 
    40 & 0.002 & 0.002 & 0.058 & 0.067 \\ \hline 
    50 & 0.001 & 0.007 & 0.040 & 0.047 \\ \hline 
    60 & 0.029 & 0.036 & 0.015 & 0.021 \\ \hline 
    70 & 0.023 & 0.015 & 0.046 & 0.051 \\ \hline 
 	\end{tabular}
    \label{table:ton_BLRmodelvalidation}
\end{table*}
\section{Conclusions}
\label{sec:ton_conclusions}



In this work, the performance of TCP NewReno has been analyzed over a random access satellite channel. Our proposed throughput estimation model, namely NewRenoSAT, accurately fits the simulation results for the satellite scenario under consideration, where losses are only due to collisions. Furthermore, our approach has been compared with other models in literature, showing that it achieves a lower estimation error.
Our simulation results support the hypothesis that collisions can be considered as random independent events on a RA channel. 
Furthermore, a simple but effective model has been provided to estimate the loss event rate $p$ and the segment loss rate $q$ at transport layer; this is the \textit{BLR model} that has shown a close match with simulation results, when used in the NewRenoSAT model.

In addition, the use of TCP enforces stability in RA channels, removing the need of the DVB-RCS2 load control mechanism, which is strongly scenario-dependent. TCP acts in a distributed way, without the need of a centralized entity in charge of enforcing potentially complex load control strategies.
The use of different waveforms and MSS sizes has been analyzed, in order to identify the combination providing the best performance in the IoT/M2M scenario under consideration, also depending on the number of RCSTs. 
Furthermore, any system configurations that lead to the fragmentation of TCP segments at MAC layer should be avoided because of the poor offered performance.
In conclusions, the use of a TCP-based application protocol MQTT-like can support a large population in a satellite system, exploiting the advantages offered by the PUB/SUB paradigm in this context.
\balance


%
\IEEEtriggeratref{0}
\bibliographystyle{IEEEtran}
\bibliography{references}

\end{document}